\documentclass[sigconf,natbib=true,anonymous=false,review=false,screen=true]{acmart}

\usepackage{acronym}
\usepackage{dsfont}
\usepackage{amsmath}

\usepackage{bm}
\usepackage{booktabs}

\usepackage[T1]{fontenc}
\usepackage{graphicx}
\usepackage{hyperref}
\usepackage{listings}
\usepackage{multirow}

\usepackage{subcaption}
\usepackage{tabularx}
\usepackage[dvipsnames]{xcolor}
\usepackage{enumerate}

\usepackage{numprint}
\usepackage{mleftright}
\usepackage{balance}

\usepackage{amsthm}
\usepackage{mathtools, nccmath, textcomp} 
\usepackage{xargs}
\usepackage{siunitx}

\usepackage{rdbdbasis}
\usepackage{rdbdanalysis}
\usepackage{rdbdcalculus}
\usepackage{rdbdmatrix}
\usepackage{rdbdstatistics}

\newcommand{\qd}{{q,d}}
\newcommand{\qdk}{{\qd,k}}

\newcommand{\posBias}{{\theta}}
\newcommand{\itemRel}{{\zeta}}

\newcommand{\posBiasBM}{{\bm{\posBias}}}
\newcommand{\itemRelBM}{{\bm{\itemRel}}}

\newcommand{\Jqd}[1][]{{J_\qd^{#1}}}
\newcommand{\sJqd}[1][]{{\tilde{J}_\qd^{#1}}}

\newcommand{\JJqd}[1][]{{\mathcal{J}_\qd^{#1}}}
\newcommand{\sJJqd}[1][]{{\tilde{\mathcal{J}}_\qd^{#1}}}

\newcommand{\Pqdk}{\posBias_k\itemRel_\qd}
\newcommandx{\sPqdk}[2][1={}, 2={}]{{\tilde{\posBias}^{#1}_k\tilde{\itemRel}^{#2}_\qd}}

\newcommandx{\LLD}[2][1={}, 2={}]{{\mathcal{L}_{\mathrm{#1}}^{#2}}}
\newcommandx{\sLLD}[2][1={}, 2={}]{{\tilde{\mathcal{L}}_{\mathrm{#1}}^{#2}}}

\newcommand{\lqdk}{{J_\qdk}}
\newcommandx{\slqdk}[1][1={}]{{\tilde{J}_{\qdk}^{#1}}}

\newcommandx{\slqd}[1][1={}]{{\tilde{J}_{\qd}^{#1}}}

\newcommand{\logL}{{\mathscr{l}}}
\newcommandx{\slogL}[1][1={}]{{\tilde{\mathscr{l}}^{{#1}}}}

\newcommand{\KAlpha}{{\bm{\alpha_\posBias}}}
\newcommand{\KBeta}{{\bm{{\beta_\posBias}}}}

\newcommand{\betaParamsPB}{{\bm{\alpha_\posBias}, \bm{\beta_\posBias}}}
\newcommand{\betaParamsQD}{{\bm{\alpha_\itemRel}, \bm{\beta_\itemRel}}}

\newcommand{\relevant}{{\mathtt{R}}}
\newcommand{\examination}{{\mathtt{E}}}
\newcommand{\click}{{\mathtt{C}}}

\newcommand{\QD}{{\mathrm{rel}}}
\newcommand{\PB}{{\mathrm{pos}}}
\newcommand{\gradQD}{{\grad_{{\QD}}}}
\newcommand{\gradPB}{{\grad_{{\PB}}}}

\newcommand{\QDsize}{{\abs{\mathcal{D}}}}

\newcommand{\LSE}{{\mathrm{LSE}}}

\newtheorem{definition}{Definition}[section]

\newcommand{\binomProbD}[4]{{\binom{#4}{#2}\qty(#1)^{#2}\qty(1-#1)^{#3}}}

\acrodef{IR}{information retrieval}
\acrodef{LTR}{learning-to-rank}
\acrodef{ARP}{average relevance position}
\acrodef{DCG}{discounted cumulative gain}
\acrodef{EM}{expectation-maximization}
\acrodef{CTR}{click-through-rate}
\acrodef{NRCTR}{normalized RCTR}
\acrodef{NDCG}{normalized DCG}
\acrodef{PL}{Plackett-Luce}
\acrodef{IPS}{inverse-propensity-scoring}
\acrodef{DR}{doubly-robust}
\acrodef{PBM}{position-based click model}
\acrodef{PGM}{probabilistic graphical model}
\acrodef{NN}{neural network}
\allowdisplaybreaks

\AtBeginDocument{%
  }

\copyrightyear{2026}
\acmYear{2026}
\setcopyright{cc}
\setcctype{by}
\acmConference[SIGIR '26]{Proceedings of the 49th International ACM SIGIR Conference on Research and Development in Information Retrieval}{July 20--24, 2026}{Melbourne, VIC, Australia}
\acmBooktitle{Proceedings of the 49th International ACM SIGIR Conference on Research and Development in Information Retrieval (SIGIR '26), July 20--24, 2026, Melbourne, VIC, Australia}
\acmDOI{10.1145/3805712.3809529}
\acmISBN{979-8-4007-2599-9/2026/07}

\begin{document}
\title{An Epistemic Position-Based Click Model: From~Interactions~to~Epistemic~Distributions~of~Relevance~and~Bias}

\author{Oscar Rolando Ramirez Milian}
\email{o.r.ramirezmilian@uva.nl}
\affiliation{%
	  \institution{University of Amsterdam}
	\city{Amsterdam}
	  \country{The Netherlands}
	}
\authornote{Work performed while at the Radboud University Nijmegen.}

\author{Harrie Oosterhuis$^{*}$}
\email{h.r.oosterhuis@uva.nl}
\affiliation{%
	\institution{University of Amsterdam}
	\city{Amsterdam}
	\country{The Netherlands}
}
\renewcommand{\authors}{Oscar Rolando Ramirez Milian and Harrie Oosterhuis}
\renewcommand{\shortauthors}{Oscar Rolando Ramirez Milian and Harrie Oosterhuis}

\begin{abstract}
User interactions with rankings are affected by both items' relevances and display positions.
Accordingly, click probabilities are often modeled as a product of relevance and position factors; and for improving recommendation and search, one needs to disentangle relevance from position bias.
However, existing click models only provide frequentist point-estimates that do not capture any measure of epistemic uncertainty.
Consequently, there is no indication of how much confidence one should have in their predictions.
In this work, we introduce the first evidential deep-learning approach to form an \emph{epistemic} alternative to the important position-based click model.
Our learned model takes as input item and position features and outputs a beta-distribution for every relevance and position-bias variable of the position-based model.
These distributions capture epistemic uncertainty about click probabilities and the underlying effects of attraction and position bias.
The main challenge of our approach is its optimization for which we propose approximation and conditioning techniques to provide numerical stability and variance reduction. Our experiments indicate that our approach captures epistemic uncertainty in predictions on previously-unseen data, whereas standard policy gradients fail to learn meaningful distributions. We believe our contribution of the first \emph{contextual epistemic click model} constitutes an important step in incorporating Bayesian uncertainty into click modeling.
\end{abstract}



\maketitle

\acresetall 

\section{Introduction}

Search and recommendation systems connect users to online content by presenting rankings of items~\citep{liu2009ltrbook, lu2012recsys}, thereby, they are vital for making  large online collections accessible and navigable~\citep{albert1999diameter}.
The ordering in these rankings heavily affect user satisfaction and what items they may consider interacting with~\citep{jarvelin2002cumulated, sanderson2010user}.
Therefore, it is important to have accurate estimates of user preferences for items to provide the best rankings; and at same time, the effect of the presented ordering makes it difficult to infer such preferences from interactions with rankings~\citep{oosterhuis-phd-thesis-2020}.
The fields of click modeling and unbiased learning-to-rank concern themselves with methods to disentangle user preferences from other factors~\citep{oosterhuis2022end, chuklin2015basic, shashank2024tutorial, joachims2017unbiased}.
The \ac{PBM}~\citep{craswell2008experimental, chuklin2015basic} is a foundational click model that remains relevant due to its combination of simplicity and effectiveness~\citep{craswell2008experimental, chen2012position, joachims2017unbiased, wang2018position, ai2018unbiased, ai2021unbiased, ovaisi2020correcting, oosterhuis2023doubly}:

\begin{definition}[\Acfi{PBM}]
\label{def:pbm}
In the \ac{PBM}, the probability of a click $(\click=1)$ on an item $d$ displayed at position $k$ of a ranking in the context $q$ is a product of the probability that position $k$ is examined $(\examination=1)$, conditioned on the position, and the probability that item is found attractive $(\relevant=1)$, conditioned on the context and item:
\begin{equation}
	\Prob{\click=1}[q,d,k]=\Prob{\examination=1}[k]\Prob{\relevant=1}[q,d]= \posBias_{k} \itemRel_{q,d} %
	.
\end{equation}

\end{definition}
The $\posBias$ and $\itemRel$ parameters are often referred to as \emph{position bias} and \emph{item relevance} (a.k.a.\ attractiveness or preference) factors respectively~\citep{craswell2008experimental, oosterhuis-phd-thesis-2020}.
We note that our notation originates from a web-search setting where $q$ is a search query and $d$ is a document~\citep{oosterhuis2023doubly};
in recommendation settings, $q$ can represent information about the user and/or the webpage at which the recommendation is made and $d$ an item to be recommended~\citep{oosterhuis-phd-thesis-2020}.
Since examination and relevance cannot be observed directly, the \ac{PBM} parameters have to be inferred from logged click data~\citep{craswell2008experimental, chuklin2015basic, wang2018position}.

Let $N_{q,d,k} \in \mathds{Z}_{\geq0}$ be the number of times item $d$ was displayed at position $k$ for query $q$ and $M_{q,d,k} \in \mathds{Z}_{\geq 0}$ the number of times $d$ was clicked at $k$ for $q$ and let $\mathbf{N}$ and $\mathbf{M}$ be tuples of all $N_{q,d,k}$ and $M_{q,d,k}$ values respectively.
Existing methods for estimating the parameters of the \ac{PBM} take one of two approaches~\citep{oosterhuis2022end}; The click model family searches for the parameters that maximize the likelihood of the data~\citep{chuklin2015basic, wang2018position, kang2025rethinking}: $
(\posBiasBM^*, \itemRelBM^*) = \arg\max_{(\posBiasBM, \itemRelBM)} \Prob{\mathbf{M}}[\mathbf{N},\posBiasBM, \itemRelBM]$. Unbiased learning-to-rank methods first estimate the examination probabilities per rank, e.g., through position randomization~\citep{wang2018position}, to use for \acl{IPS} estimates of relevance~\citep{joachims2017unbiased, wang2016learning}, for example, with estimated propensities~$\rho$: $
 \hat{\itemRel}^\text{IPS}_{q,d}= \frac{1}{\sum_{k=1}^K N_{q,d,k}} \sum_{k=1}^K \frac{M_{q,d,k}}{\rho[\examination = 1 \mid k]}
$. A significant limitation of these approaches is that they only provide pointwise estimates for the parameters~\citep{hullermeier2021aleatoric}.
Therefore, they do not give any indication of how reliable their estimates are, nor how much confidence one should have in their value~\citep{armen2009aleatory}.
Moreover, as a result, they also do not provide insight in the possible risks in using them for downstream tasks.
This is especially concerning since the \ac{PBM} explains observations by products of probabilities, and consequently, generally many different parameter values can explain observed click data equally well~\citep{oosterhuis2022end, hager2025unidentified}.
This is part of a larger limitation of the click modeling and unbiased learning-to-rank fields, as --to the best of our knowledge-- all state-of-the-art methodologies only provide pointwise estimates~\citep{joachims2017unbiased, shashank2024tutorial, oosterhuis-phd-thesis-2020, oosterhuis2020unbiased, kang2025rethinking}.

This work proposes the first \emph{contextual epistemic parameter estimation method} for the \ac{PBM}; our approach is based on evidential deep-learning~\citep{sensoy2018evidential, soleimany2021evidential, malinin2018predictive, malinin2019reverse, amini2020deep} and learns a distribution for each of its parameters that represent epistemic uncertainty.

Our model consists of a Beta distribution for each parameter, i.e., $\theta_k  \in (0,1)$ for every $k$ and $\itemRel_{q,d}  \in (0,1)$ for every $(q,d)$, which are treated as independent random variables.
Accordingly, we optimize the model by searching for the best $\alpha$ and $\beta$ parameters of the Beta distributions, i.e., $(\alpha_k, \beta_k)$ for every $\theta_k$, and $(\alpha_{q,d}, \beta_{q,d})$ for every $\itemRel_{q,d}$, by maximizing the likelihood of the data under the $\alpha$ and $\beta$ parameters; with $\posBiasBM \in (0,1)^K$, $\itemRelBM \in (0,1)^{QD}$, and their distribution parameters $\bm{\alpha}_{\posBiasBM}, \bm{\beta}_{\posBiasBM}, \bm{\alpha}_{\itemRelBM}, \bm{\beta}_{\itemRelBM}$ as tuples of all $\alpha_k$, $\beta_k$, $\alpha_{q,d}$ and $\beta_{q,d}$ respectively:

\begin{equation}
\arg\hspace{-1em}\max_{\hspace{-1.5em}(\bm{\alpha}_{\posBiasBM}, \bm{\beta}_{\posBiasBM}, \bm{\alpha}_{\itemRelBM}, \bm{\beta}_{\itemRelBM})}
\int
\probR{\mathbf{M}}[\mathbf{N}, \itemRelBM,\posBiasBM]
\probD{\itemRelBM, \posBiasBM}[\bm{\alpha}_{\itemRelBM}, \bm{\beta}_{\itemRelBM},\bm{\alpha}_{\posBiasBM}, \bm{\beta}_{\posBiasBM}]
d \itemRelBM d \posBiasBM
.
\label{eq:intro:doubleintegral}
\end{equation}
Because the number of queries and items ($QD$) is large, we optimize a neural network to predict $\alpha_{q,d}$ and $\beta_{q,d}$ based on features $X_{q,d}$.
Vice versa, because the number of positions ($K$) is very limited, we learn $\alpha_k$ and $\beta_k$ values directly.
However, a direct application of Monte Carlo integration to solve (\ref{eq:intro:doubleintegral}) poses substantial challenges. In particular, the numerical precision of the sample values and the high variance inherent in the gradient estimation are main difficulties.
As a solution, we introduce a novel approach that optimizes a numerically-stable approximation of the log-likelihood and reduces variance through conditioning on partial samples.

Our novel method is the first \emph{contextual epistemic click model}, and provides a foundation for future click models that use evidential deep learning for uncertainty quantification~\citep{sensoy2018evidential, malinin2018predictive, amini2020deep, pandey2023evidential, knyazev2023lightweight}.

\section{Related Work}

\textbf{Click models.}\;
Early click models were \acp{PGM} handcrafted for the classic web search setting~\citep{chuklin2015basic, craswell2008experimental, richardson2007clicks, dupret2008userbrowsing, Liu2024}; among them is the \ac{PBM}~\citep{craswell2008experimental, richardson2007clicks}. 
Later, more complex click models consider sequences of actions, such as the cascading click model~\citep{craswell2008experimental} or the user browsing model~\citep{dupret2008userbrowsing}.
\Ac{NN} click models were introduced to avoid the assumptions inherent in \ac{PGM} design and their constraints~\citep{borisov2016neural, borisov2018clicksequence, yu2019rankbiased}.
Many click models have been designed with specific settings or applications in mind, e.g., for specific interfaces such as grid layouts for image search and carousel interfaces for recommendation systems~\citep{kang2025rethinking, rahdari2024towards, rahdari2022magic, xie2018image, zhuang2021crosspositional, chierichetti2011twodimensional, guo2020debiasing}.
In the unbiased learning-to-rank field, click models are often applied to separate the effect of user preference from examination behavior~\citep{hager2023contrast, wang2016learning, wang2018position, guo2020debiasing}.
This has lead to \ac{NN} click models that apply a two-tower structure, where the model is divided into two parts with either position or item features~\citep{hager2025unidentified, zhuang2021crosspositional, yan2022revisiting}.
The many \acl{IPS} methods in the field are commonly build on the assumption that user clicks come from a \ac{PBM} click model and require accurate probability examinations~\citep{wang2016learning, joachims2017unbiased, wang2018position, ai2018unbiased, ovaisi2020correcting}.
Although, a few works also assume trust-bias or cascading click models~\citep{oosterhuis2023doubly, vardasbi2020inverse, vardasbi2020cascade, kiyohara2022cascade, agarwal2019addressing}.

As a result, methods for the estimation of \ac{PBM} parameters remain important to the unbiased off-policy evaluation and optimization of rankings~\citep{hager2023contrast, chen2012position, joachims2017unbiased, wang2018position, ai2018unbiased, ai2021unbiased, ovaisi2020correcting, oosterhuis2023doubly}.

\citet{Liu2009bbm} also proposed a Bayesian click model, in contrast with our approach, their model is \emph{non-contextual};
They do not use document features, but instead optimize model parameters in a lookup table, consequently, they cannot handle previously-unseen or long-tail documents for which there is little to no data.
Several categorizations of click models have been made:
\citet{chuklin2015basic} give an overview of \ac{PGM} click models, and \citet{Liu2024} give two separate categorizations for  \ac{PGM} and  \ac{NN} click models.
Recently, \citet{kang2025rethinking} propose a theory-based categorization that enables the comparison of click models of different types and for different interfaces.
Together, they provide a comprehensive overview of existing click models from which it appears that, besides \citet{Liu2009bbm},  no other previous work has considered Bayesian click models.

\noindent
\textbf{Uncertainty estimation and evidential learning.}\;
Uncertainty is a fundamental concept in machine learning, generally, a distinction is made between \emph{aleatoric} uncertainty stemming from randomness inherent in an outcome to be predicted; and \emph{epistemic} uncertainty  stemming from a lack of knowledge~\citep{armen2009aleatory}.
There are numerous approaches to uncertainty estimation; \citet{hullermeier2021aleatoric} provide a comprehensive overview.
The most relevant to this work is the \emph{evidential deep learning} methodology; originating from the work by \citet{malinin2018predictive} who proposed \emph{prior networks} for evidential regression, and \citet{sensoy2018evidential} who proposed \emph{evidential classification}.
The core idea is to let a network predict the parameters of a distribution over the outcome domain, thereby representing the epistemic uncertainty of a prediction~\citep{sensoy2018evidential, malinin2018predictive, hullermeier2021aleatoric, knyazev2023lightweight}.
Whilst the evidential approach has become popular~\citep{soleimany2021evidential, malinin2018predictive, malinin2019reverse, amini2020deep}, likely due to its easy applicability, some have questioned whether it accurately learns to capture epistemic uncertainty~\citep{bengs2022pitfalls, pmlr-v235-juergens24a}.
In response, \citet{meinert2023unreasonable} argue that evidential deep learning is a useful heuristic that gives a reasonable proxy of epistemic uncertainty; we discuss this further in Section~\ref{sec:background:heuristic}.

\begin{figure*}[t]
	\centering
	{
		\renewcommand{\arraystretch}{0.5}
		\setlength\tabcolsep{0.01pt}
		\begin{tabular}{c c |c| c c}
			\includegraphics[scale=0.325]{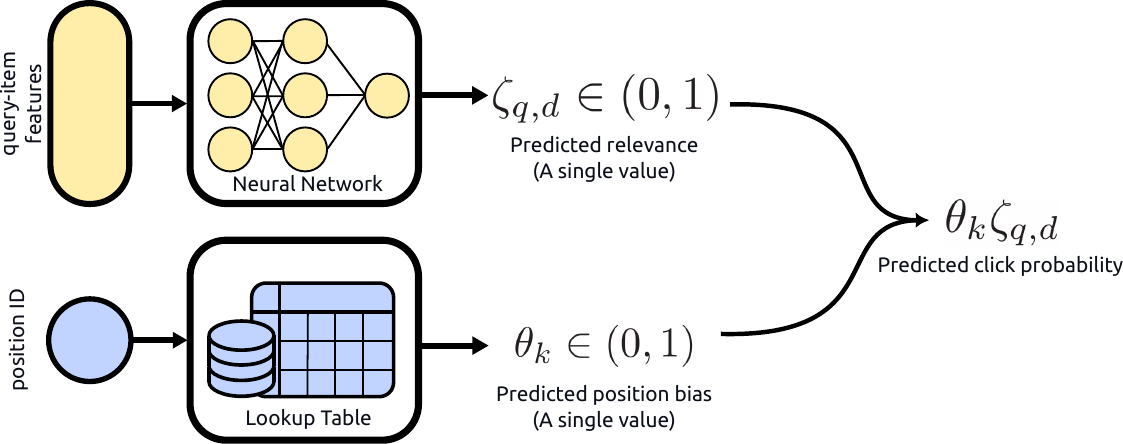} 
			&
			\hspace{0.95em}
			&
			&
			\hspace{0.95em}
			&
			\includegraphics[scale=0.325]{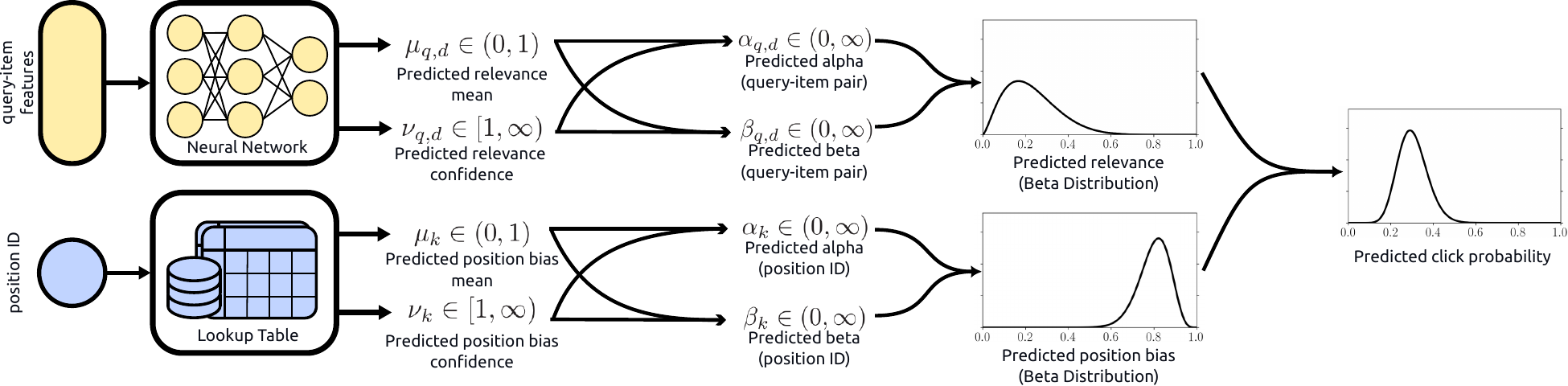} \\
			&\hspace{1em}& & \hspace{1em} & \\
			(a) Pointwise deep learning.&\hspace{1em}& & \hspace{1em} & (b) Evidential deep learning.
		\end{tabular}
	}
	\vspace{-0.5em}
	\caption{Schematic comparison of the two deep learning instantiations of a position-based click model.}
	\label{fig: PBM vs EPBM}
	\vspace{-0.5em}
\end{figure*}

\section{Preliminaries: Notation and Epistemic PBM}

In our setting, the observed data are query-item pairs, their displays to the users and the clicks received from displays.
We have the query-item pairs $(q,d) \in \mathcal{D}$ with feature-vector representations $X_{q,d}$.
For every position $k \in \{1,2,\ldots,K\}$ and pair $(q,d)$, $N_{q,d,k} \in \mathds{Z}_{\geq 0}$ indicates the number of times the item $d$ was displayed at position $k$ for query $q$,
and $M_{q,d,k}$ the number of times it was clicked.
For convenience, we also define $W_{q,d,k} = N_{q,d,k} - M_{q,d,k}$ as the number of displays without clicks. Lastly, we assume that there are many more pairs than display positions: $|\mathcal{D}| >> K$, which is generally true for recommendation and retrieval~\citep{albert1999diameter}.

In the \ac{PBM}, the probability of a click is a product of a position factor $\theta_k$ and a relevance factor $\zeta_{q,d}$ (Definition~\ref{def:pbm}).
We assume that all clicks are independent Bernoulli variables, an assumption that often goes implicitly with the \ac{PBM}~\citep{chuklin2015basic}.
Therefore, each sum of clicks $M_{q,d,k}$ is binomially distributed when conditioned on $N_{q,d,k}$:
\begin{equation}
	\probR{M_\qdk \big| N_\qdk, \posBias_k, \itemRel_\qd} 
	\!=\!
	\binom{N_\qdk}{M_\qdk}\!\qty(\Pqdk)^{\!M_\qdk}\!\qty(1\!-\!\Pqdk)^{\!W_\qdk}\!\!.
	\label{eq:binomprob}
\end{equation}
To keep our notation brief, we define $\boldsymbol{\theta} = \qty(\theta_k : k\in (1,2,\ldots,K) )$  as the sequence of all $\theta_{k}$ and define $\boldsymbol{\zeta}$ analogously.
Thereby, $\boldsymbol{\theta}$ and $\boldsymbol{\zeta}$ capture the \emph{aleatoric} uncertainty of our model~\citep{hullermeier2021aleatoric}, as they aim to describe the stochasticity inherent in the clicking behavior of users.

To capture \emph{epistemic} uncertainty, we opt for an evidential approach~\citep{liu2001propositional,malinin2018predictive, amini2020deep} by fitting a parametric distribution over the possible values of $\boldsymbol{\theta}$ and $\boldsymbol{\zeta}$.
Specifically, we rely on Beta distributions as each individual $\theta$ or $\zeta$ variable is in the interval $(0,1)$.
Accordingly, we model each $\theta_k$ as a sample from a Beta distribution specific to position $k$: $\posBias_k\sim\BetaD{\alpha_k}{\beta_k}$; and analogously: $\itemRel_\qd\sim\BetaD{\alpha_\qd}{\beta_\qd}$. 
We model the joint probability density function as the product of the individual marginal densities:
\begin{equation}
	\probD{\posBiasBM, \itemRelBM}[\boldsymbol{\alpha_\posBias, \beta_\posBias, \alpha_\itemRel, \beta_\itemRel}]=\probD{\posBiasBM}[\boldsymbol{\alpha_\posBias, \beta_\posBias}]\probD{\itemRelBM|\boldsymbol{\alpha_\itemRel, \beta_\itemRel}}.
\end{equation}
This implies that the $\theta_k$ and $\itemRel_{\qd}$ random variables are independent, i.e., conditioning on one does not change the others' distributions:
\begin{equation}
\begin{split}
	\probD{\posBiasBM}[\boldsymbol{\alpha_\posBias, \beta_\posBias}]&= \prod_{k=1}^K\probD{\posBias_k|\alpha_k, \beta_k},\\
	\probD{\itemRelBM}[\boldsymbol{\alpha_\itemRel, \beta_\itemRel}]&= \prod_{\qd \in \mathcal{D}}\probD{\itemRel_\qd|\alpha_\qd, \beta_\qd}.
\end{split}
\end{equation}
Again for brevity, we define $\boldsymbol{\alpha_\posBias} = \qty(\alpha_k : k\in (1,2,\ldots,K) )$, $\boldsymbol{\alpha_\itemRel} = \big(\alpha_{q,d} : (q,d)\in\mathcal{D}\big)$ and analogously $\boldsymbol{\beta_\posBias}$ and $\boldsymbol{\beta_\itemRel}$.
Thus, $\boldsymbol{\alpha_\posBias}$, $\boldsymbol{\beta_\posBias}$, $\boldsymbol{\alpha_\itemRel}$ and $\boldsymbol{\beta_\itemRel}$ describe the \emph{epistemic} uncertainty of our model, i.e., how probable each possible value of $\boldsymbol{\theta}$ and $\boldsymbol{\zeta}$ is predicted to be.
To the best of our knowledge, this is the first epistemic \ac{PBM} and potentially the first \emph{epistemic click model} altogether~\citep{chuklin2015basic, Liu2024, kang2025rethinking}.

Finally, the question now becomes how to find the values for $\boldsymbol{\alpha_\posBias}$, $\boldsymbol{\beta_\posBias}$, $\boldsymbol{\alpha_\itemRel}$ and $\boldsymbol{\beta_\itemRel}$ such that our epistemic model best predicts the true $\posBiasBM$ and $\itemRelBM$ values and an appropriate uncertainty about its predictions.
Following evidential deep learning methodology~\citep{malinin2020regression, amini2020deep}, we search for a model that maximizes the likelihood of the observed data:
\begin{equation}
\begin{split}	
	\mathcal{L}&=
	\!\!\int
	\probR{\mathbf{M}}[\mathbf{N}, \itemRelBM,\posBiasBM]
	\probD{\itemRelBM}[\bm{\alpha}_{\itemRelBM}, \bm{\beta}_{\itemRelBM}]
	\probD{\posBiasBM}[\bm{\alpha}_{\posBiasBM}, \bm{\beta}_{\posBiasBM}]
	d \itemRelBM d \posBiasBM \\[-1.3ex]
	&=\mathds{E}\Big[ \prod_{\qd, k}\probR{M_\qdk}[N_\qdk, \posBias_k, \itemRel_\qd] \, \Big|\, \betaParamsQD, \betaParamsPB \Big]
	.
	\label{eq: Likelihood}
\end{split}
\end{equation}
Accordingly, our goal is to develop a method for optimizing $\boldsymbol{\alpha_\posBias}$, $\boldsymbol{\beta_\posBias}$, $\boldsymbol{\alpha_\itemRel}$ and $\boldsymbol{\beta_\itemRel}$ to maximize the likelihood of our observed data.
Because the number of positions $K$ is limited, we learn $\boldsymbol{\alpha_\posBias}$ and $\boldsymbol{\beta_\posBias}$ directly (in a lookup table).
Conversely, since the number of query-item pairs is enormous, we learn a neural network to predict $\alpha_\qd$ and $\beta_\qd$ for a given feature vector $X_{\qd}$, thereby making our approach generalizable to previously-unseen queries and items.

\section{Background}

\subsection{Na\"ive Monte Carlo gradient estimation}
\label{sec: Naive Monte Carlo Gradient Estimation}
Since the likelihood is an expectation (\ref{eq: Likelihood}), it can be estimated unbiasedly through a straightforward Monte Carlo estimation~\citep{hammersley2013monte}. We can take $S$ samples for each of the $\posBias$ and $\itemRel$ parameters according to $\betaParamsPB$ and $\betaParamsQD$, which we group in vectors for convenience:
\begin{align}
\tilde{\posBias}_k^{(i)} \sim \BetaD{\alpha_k}{\beta_k}, \qquad\quad\, &\;
\bm{\tilde{\posBias}}^{(i)} = \big(\tilde{\posBias}_1^{(i)}, \dots, \tilde{\posBias}_K^{(i)}\big),
\\[-1ex]
\tilde{\itemRel}_\qd^{(i)} \sim \BetaD{\alpha_\qd}{\beta_\qd}, \qquad &\;
\bm{\tilde{\itemRel}}^{(i)} = \big(\tilde{\itemRel}_\qd^{(i)} : (\qd)  \in  \mathcal{D}\big).
\end{align}
We can evaluate the probability of our observations conditioned a single sample of  position bias $\tilde{\posBiasBM}^{(i)}$ and relevance values $\tilde{\itemRelBM}^{(i)}$:
\begin{equation}
	\tilde{L}\big(\bm{\tilde\itemRel}^{(i)}\!\!\!, \bm{\tilde\posBias}^{(i)}\big) \!\! \triangleq  \! \tilde{L}^{(i)}  \!\! \triangleq  \!\!
	\prod_{\qd, k}\!\!\binomProbD{\sPqdk[(i)][(i)]}{\!M_{\qd, k}\!}{\!W_{\qd, k}\!}{N_{\qd, k}}
	\!\!\!\!\!.
	\label{eq:Pointwise Likelihood}
\end{equation}
The mean value of $\tilde{L}^{(i)}$ provides an unbiased estimate of the likelihood of our model, additionally,  the gradient of the likelihood can be unbiasedly estimated with the log-derivative trick~\citep{williams1992simple}:

\begin{equation}	
	\begin{split}
	\tilde{\mathcal{L}}
	\approx
	\frac{1}{S}\sum_{i=1}^{S} \tilde L ^{(i)},
	\quad
		\grad \mathcal{L}
		\approx
		\frac{1}{S}\sum_{i=1}^{S} \tilde L ^{(i)}\mleft(\grad\log \probD{\bm{\tilde\itemRel}^{(i)}|\betaParamsQD}\mright.\;\;
		\\[-2ex]
		+ \mleft.\grad\log \probD{\bm{\tilde\posBias}^{(i)}|\betaParamsPB} \mright). 
		\label{eq: Monte Carlo Gradient Estimation}
	\end{split} 
\end{equation}
A lower-variance variation of this estimate can be made with leave-one-out baseline-corrections~\citep{kool2019buy}:
\begin{equation}
\begin{split}	
	&\grad \mathcal{L}
	\approx
	\frac{1}{S}\sum_{i=1}^{S}
	\mleft(
	\mleft(
	 \tilde{L}^{(i)}
	 - 
	 \frac{1}{S-1}\sum_{j=1}^{S} \mathds{1}[j \not = i] \tilde{L}^{(j)}
	 \mright)\mright. \, \times
	 \\[-0.8ex] &\qquad\quad\;
	 \mleft.\mleft( \grad\log \probD{\bm{\tilde\itemRel}^{(i)}|\betaParamsQD}  + \grad\log \probD{\bm{\tilde\posBias}^{(i)}|\betaParamsPB} \mright)\mright).
\label{eq: Monte Carlo Gradient Estimation-Baseline}
\end{split}
\end{equation}
Nevertheless, due to the high dimensionality of $\bm{\tilde\itemRel}$ and $\bm{\tilde\posBias}$, the variance of this estimate is impractically high, and the values of $\tilde{L}^{(i)}$ are incredibly small, leading to numerical precisions problems when estimating their mean or the estimated policy gradient.

\subsection{Evidential deep learning as a heuristic}
\label{sec:background:heuristic}
Despite the popularity of evidential deep learning due to its relative practicality~\citep{malinin2018predictive, amini2020deep, sensoy2018evidential, pandey2023evidential}, there has also been criticism. 
Namely, \citet{pmlr-v235-juergens24a} criticise \emph{second-order learners} which predict probability distributions over outcomes and are optimized with standard loss functions. \citet{bengs2022pitfalls} and \citet{pmlr-v235-juergens24a} provide proofs that the distributions that minimize evidential deep learning losses are Dirac delta functions that concentrate all probability density on a single point.
Therefore, loss minimization is not expected to properly calibrate the shape of predicted distributions, and thus, does not appropriately represent  epistemic uncertainty. 

While their conditions do not entirely apply to our setting, a similar behavior can be found for our approach. Let $\bm{\itemRel}^*$ and $\bm{\posBias}^*$ be values that maximize the likelihood of our observations:
\begin{equation}
 L(\itemRelBM^*, \posBiasBM^*)=\sup_{\itemRelBM, \posBiasBM} L(\itemRelBM, \posBiasBM)=\max_{\itemRelBM, \posBiasBM} L(\itemRelBM, \posBiasBM)\geq L(\itemRelBM,\posBiasBM).
\end{equation}
The value of $L$ evaluated at any $\itemRelBM^*, \posBiasBM^*$ of the maximizing set, is always an upper bound for $\mathcal{L}$, for all $\betaParamsQD$ and  $\betaParamsPB$:
\begin{equation}
	L(\itemRelBM^*, \posBiasBM^*)\geq \mathds{E}\big[L(\itemRelBM, \posBiasBM) \big|\betaParamsQD, \betaParamsPB \big]=\mathcal{L}.
\end{equation}
A sequence of parameters $\big(\bm{\alpha}^{(n)}_\itemRelBM\!\!\!,\bm{\beta}^{(n)}_\itemRelBM\big)_n$ and $\big(\bm{\alpha}^{(n)}_\posBiasBM\!\!\!,\bm{\beta}^{(n)}_\posBiasBM\big)_n$ can always be constructed such that $\alpha_\qd^{(n)}\!\! + \beta_\qd^{(n)}\!\! \to \!\infty$, $\alpha_k^{(n)}\!\! +  \beta_k^{(n)}\!\! \to \!\infty$ as $n\! \to \! \infty$ and the corresponding Beta distributions concentrate  probability mass monotonically around 
$\itemRelBM^*$ and $\posBiasBM^*$.
In other words, there is always a sequence of Beta distributions whose probability measures converge weakly to Dirac measures supported on $\itemRelBM^*$ and $\posBiasBM^*$:
\begin{equation}	
	\BetaD{\alpha^{(n)}_\qd}{\beta^{(n)}_\qd}\xrightarrow{w}\delta_{\itemRel^*_\qd},
	\quad
	\BetaD{\alpha^{(n)}_k}{\beta^{(n)}_k}\xrightarrow{w}\delta_{\posBias^*_k}.
\end{equation}
The Portmanteau theorem ~\citep{billingsley2013convergence} implies that $\mathcal{L}$, evaluated as an expectation of $L$ under such distributions, tends to the value:
\begin{equation}	
	L\big(\itemRelBM^*, \posBiasBM^*\big)=\textstyle\EExpectation{L(\itemRelBM,\posBiasBM)}[\itemRelBM\sim\delta_{\itemRelBM^*}, \posBiasBM\sim\delta_{\posBiasBM^*}].
\end{equation}

In summary, any Dirac measure supported on a maximizer of $L$ attains the maximum of $\mathcal{L}$.
Consequently, in our setting the optimizer can asymptotically concentrate all the probability mass on singleton sets, by choosing Beta distributions with sufficiently large parameters.
Thus, in theory, our optimization does not converge on an appropriate shape of our epistemic distribution, analogous to the criticism of \citet{pmlr-v235-juergens24a} and \citet{bengs2022pitfalls}.

\label{sec:background:entropy}
Nevertheless, \citet{meinert2023unreasonable} argue that practitioners should not treat evidential deep learning as a method for exact epistemic uncertainty quantification, but instead as a useful heuristic that provides a reasonable proxy of uncertainty.
This could explain its continued popularity~\citep{malinin2018predictive, amini2020deep, sensoy2018evidential, pandey2023evidential}.
A common heuristic for evidential deep learning is the usage of entropy regularization.
For our setting, let $\mathbf{H}$ be the entropy of the Beta distributions: 
\begin{equation}
	\mathbf{H}(\betaParamsQD, \betaParamsPB)=\sum_{\qd}H(\alpha_\qd,\beta_\qd)+\sum_k H(\alpha_k, \beta_k),
	\label{eq:background:entropy}
\end{equation}
where $H(\alpha, \beta)$ is the entropy of an individual beta distribution: 
\begin{equation}
	\begin{split}
		H(\alpha,\beta) = \log(\mathrm{B}\qty(\alpha, \beta))-(\alpha-1)\psi(\alpha)-(\beta-1)\psi(\beta)\phantom{.}&\\
			+\,(\alpha+\beta-2)\psi(\alpha+\beta).&
	\end{split}
	\label{eq:background:entropybeta}
\end{equation}
Entropy regularization is applied with the weight $\lambda \in \mathds{R}_{\geq 0}$:
\begin{equation}
\mathcal{L}' = \mathcal{L} + \lambda\mathbf{H}(\betaParamsQD, \betaParamsPB).
\end{equation}
In practice, the entropy regularization incentives the epistemic model to spread its confidence over multiple hypotheses.
In other words, if there are multiple \ac{PBM} parameters that result in comparable likelihood, then the epistemic model is incentivized to spread its probability mass over all of them.
Together with early stopping, this prevents the epistemic model to collapse into a pointwise model, and can result in a useful proxy of uncertainty in practice~\citep{meinert2023unreasonable}.

\section{Method: Optimizing the Epistemic PBM}
\label{sec:method}

This section introduces our novel methodology for estimating and optimizing the likelihood of our epistemic \ac{PBM}.
The main challenge of estimating (\ref{eq: Monte Carlo Gradient Estimation}) and (\ref{eq: Monte Carlo Gradient Estimation-Baseline}) is that it involves averaging over incredibly small values, this results in numerical precisions issues and exuberates variance-related problems during optimization.
We propose \emph{numerically stable computation}, \emph{conditioning on partial samples} and \emph{self-normalization} to overcome this challenge.

\subsection{Numerical stability for loss estimation}
\label{sec:method:LSE}
As they represent the probability of a large number of observations, the values $\tilde{L}^{(i)}<<1$ can become extremely small.
Therefore, na\"ively taking the mean of these values in (\ref{eq: Monte Carlo Gradient Estimation}) can result in underflow errors during computation.
A natural solution is to use their log values to estimate the log-likelihood instead:
\begin{equation}
	\logL(\posBiasBM, \itemRelBM)\triangleq\log(L(\posBiasBM, \itemRelBM)).
\end{equation}
Accordingly, the mean in (\ref{eq: Monte Carlo Gradient Estimation}) can done in the log space with the log-values $\slogL[(i)]$ of the likelihood samples $\tilde L^{(i)}$:
\begin{equation}
	\log\!\big(\tilde{\mathcal{L}}\big) =\log\!\Big(\frac{1}{S} \sum_{i=1}^{S}\e^{\slogL[(i)]}\Big)=\log\!\Big(\sum_{i=1}^{S}\e^{\slogL[(i)]}\Big)-\log(S).
\end{equation}
The log-sum-exp operation~\citep{blanchard2021accurately}, $\LSE:\RealS^S\!\to\!\RealS$, is commonly used to avoid underflow when operating very small numbers.
For numerically stability, it translates the log values at the start and end of the operations, while staying mathematically equivalent:
\begin{equation}
	\LSE(z_1,\ldots, z_S) \triangleq \max(z_1,\ldots, z_S) +\log\!\Big(\!\sum_{i=1}^{S}\e^{z_i- \max(z_1,\ldots, z_S)} \!\Big).
\end{equation}
The $\LSE$ can thus give a more stable estimate of the log-likelihood:
\begin{equation}
	\log\!\big(\tilde{\mathcal{L}}\big)  = \LSE\big(\slogL[(1)],\ldots,\slogL[(S)]\big)-\log(S). \label{eq: Vanilla Estimator}
\end{equation}
We note that while $\LSE$ improves stability, it is not immune to numerical precision errors.
In particular, if one of the samples $\slogL[(i)]$ dominates the others by being much larger: $\forall j \not= i, \, \slogL[(i)] >> \slogL[(j)]$, then $\LSE(\slogL[(1)],\ldots,\slogL[(S)]) \approx \max(\slogL[(1)],\ldots,\slogL[(S)]) = \slogL[(i)]$. Since:
\begin{equation}
	\max\big(\slogL[(1)],\ldots,\slogL[(S)]\big)-\log\!\big(S\big) \leq \log\!\big(\sLLD\big),
\end{equation}
we can expect an underestimate of the log-likelihood.
A key insight is that this issue is less likely if the values over which the $\LSE$ is computed are closer to each other.

\subsection{Conditioning on position bias}
\label{sec:method:conditioning}
In order to have $\slogL$ values closer to each other, we search for a lower variance estimator.
Our first insight is that (\ref{eq: Monte Carlo Gradient Estimation}) uses samples of a combination of position bias $\posBiasBM$ and relevance $\itemRelBM$ values, which are independent in our epistemic \ac{PBM}.
Therefore, the variance of the na\"ive estimator (\ref{eq: Monte Carlo Gradient Estimation}) comes from both variable sets, and thus, an estimator conditioned on one of the two could result in lower variance.
Since we assumed that there are much fewer positions than query-item pairs ($K << |\mathcal{D}|$), we condition on position bias.

For brevity, we use $\lqdk$ to denote the probability of $M_\qdk$ clicks given $N_\qdk$ displays and specific values for $\posBias_k$ and $\itemRel_\qd$ (cf.\ (\ref{eq:binomprob})):
\begin{equation}
	\lqdk(\posBias_k,\itemRel_{\qd}) \triangleq \lqdk \triangleq \probR{M_\qdk|N_\qdk, \posBias_k, \itemRel_\qd}.
\end{equation}
Similarly, the probability of the observed clicks at every position for a single item-pair with comparable conditionals is:
\begin{equation}
	\Jqd\big(\posBiasBM, \itemRel_{\qd}\big) \triangleq \Jqd \triangleq \prod_k \lqdk \big(\posBias_k, \itemRel_\qd\big).
\end{equation}
We proceed by replacing the conditional on the specific values of the relevance $\itemRel_\qd$ with our epistemic parameters $\alpha_\qd$, $\beta_\qd$, however, we keep the condition on specific values for the position bias $\posBiasBM$:
\begin{equation}
	\JJqd\big(\posBiasBM,\alpha_\qd,\beta_\qd\big) \triangleq \JJqd \triangleq \EExpectation{\Jqd\big(\posBiasBM, \itemRel_{\qd}\big)\big|\posBiasBM}[\itemRel_\qd].
\end{equation}
Importantly, through the conditioning on $\posBiasBM$, every $\JJqd$ is independent of each other.
We rewrite the likelihood (\ref{eq: Likelihood}) as: %
\begin{equation}
	\mathcal{L}
	= \mathop{\mathds{E}}_{\posBiasBM, \itemRelBM}\Big[ \prod_{\qd, k}\lqdk \Big]
	= \mathop{\mathds{E}}_{\posBiasBM}\Big[\mathop{\mathds{E}}_{\itemRelBM}\Big[ \prod_{\qd, k}\lqdk  \,\Big|\, \posBiasBM \Big]\Big],
\end{equation}
using the independence of $\JJqd$ when conditioned on $\posBiasBM$:
\begin{equation}
	\mathcal{L}
	= \mathop{\mathds{E}}_{\posBiasBM}\Big[\prod_{\qd} \mathop{\mathds{E}}_{\itemRelBM_{\qd}}\Big[ \Jqd  \,\Big|\, \posBiasBM \Big]\Big]
	= \mathop{\mathds{E}}_{\posBiasBM}\Big[\prod_{\qd}\JJqd \Big]
	.
	\label{eq:method:conditionproduct}
\end{equation}
The variance of an expectation conditioned on a variable is guaranteed to be smaller or equal than the total variance:
\begin{equation}
        \mathop{\mathds{V}}_{\posBiasBM} \Big[ 
        \mathop{\mathds{E}}_{\itemRelBM}\Big[ \prod_{\qd} \Jqd  \,\Big|\, \posBiasBM \Big]
        \Big]
	\leq
	\mathop{\mathds{V}}_{\posBiasBM,\itemRelBM} \Big[ \prod_{\qd} \Jqd  \Big]
	.
	\label{eq: Conditional variance inequality.}
\end{equation}
Thus, a potential estimator could use intermediate conditional estimates to lower variance, and thus, improve computational stability.

To construct such an estimator, we first divide our sampling procedure to separately take $S_\PB$ samples $\posBiasBM^{(i)}\sim\BetaD{\KAlpha}{\KBeta}$ and $S_\QD$ samples $\itemRel^{(j)}_\qd \sim \BetaD{\alpha_\qd}{\beta_\qd}$ for every query-item pair.
Subsequently, we can estimate $\JJqd$ using a single $\posBiasBM^{(i)}$ sample while averaging over the $S_\QD$ $\itemRel^{(j)}_\qd$ samples:
\begin{equation}
\sJJqd\big(\posBiasBM^{(i)}, \alpha_\qd, \beta_\qd\big)
\triangleq \sJJqd[(i)] \triangleq 
\frac{1}{S_\QD}\sum_{j=1}^{S_\QD}\sJqd\big(\posBiasBM^{(i)},\itemRel_{\qd}^{(j)}\big)
\!\approx\!
\JJqd.
\label{eq:method:intermediate}
\end{equation}

Based on (\ref{eq:method:conditionproduct}), we estimate the log-likelihood with $\LSE$: 
\begin{equation}
	\log\!\big(\tilde{\mathcal{L}}\big)\! \approx \LSE \Big(\!\sum_{\qd}\!\log(\sJJqd[(1)]\!),\ldots,\!\!\sum_{\qd}\!\log(\sJJqd[(S_\PB)]\!)\!\Big)\!-\!\log(S_\PB).
\label{eq:method:condition}
\end{equation}
Importantly, the intermediate estimator (\ref{eq:method:intermediate}) only concerns an expectation over a single random variable $\itemRel_\qd$, and (given the intermediate estimates) the log-likelihood estimator (\ref{eq:method:condition}) only concerns an expectation over the $K$ random variables in $\posBiasBM$.
Thereby, the variance of both these estimators is lower than when considering samples of all variables at once (cf.\ (\ref{eq: Conditional variance inequality.})), resulting in a better sample-efficiency.
Additionally, this brings the values in the $\LSE$ operations closer to each other, thus resulting in more stable computations as well.

\subsection{Gradient estimation with self-normalization}
The previous subsections concerned the estimation of the likelihood, however, for optimization a stable estimate of its gradient is equally important.
The gradient estimator in (\ref{eq: Monte Carlo Gradient Estimation}) multiplies likelihood values conditioned on single samples of parameter values $\tilde{L}^{(i)}$ with a gradient w.r.t.\ the log-probabilities of the samples.
However, generally, the likelihood values are extremely close to zero $\tilde{L}^{(i)} \approx 0$ resulting in a near-zero gradient, leading to unstable optimization.

In response, we note that for optimization a gradient's direction is generally more important than its magnitude.
Inspired by self-normalized importance sampling~\citep{swaminathan2015selfnormalized, cardoso2022snis}, we apply self-normalization by dividing the $\tilde{L}^{(i)}$ values by the estimated overall likelihood $\tilde{\mathcal{L}}$; we note that this is equivalent to optimizing the log-likelihood since:
\begin{equation}
	\grad\log(\mathcal{L}) =\frac{\grad\mathcal{L}}{\mathcal{L}}. \label{eq: grad-log}
\end{equation}
Importantly, this maintains the direction of the gradient whilst also normalizing the small values for $\tilde{L}^{(i)}$\!.
Furthermore, self-normalization can be computed more stably with a softmax $\bm{\sigma}$ using $\LSE$:
\begin{equation}
\bm{\sigma}_i\qty(z_1,\cdots,z_S) \triangleq \bm{\sigma}(z_i)
\triangleq
\frac{\e^{z_i}}{ \sum_{j=1}^{S}\e^{z_j}}
=
\e^{z_i-\LSE(z_1,\ldots, z_S)}.
\label{eq: softmax}
\end{equation}
For the naive estimator (\ref{eq: Monte Carlo Gradient Estimation}), self-normalization results in:
\begin{align}
&\grad\log(\sLLD)   \label{eq: grad Vanilla} \\
	&\approx  \frac{1}{S}\sum_{i=1}^{S}\frac{L^{(i)}}{\sLLD} \mleft(\grad\log \probD{\bm{\tilde\itemRel}^{(i)}|\betaParamsQD} + \grad\log \probD{\bm{\tilde\posBias}^{(i)}|\betaParamsPB} \mright)
	\nonumber \\
	&=\frac{1}{S}\sum_{i=1}^{S}\bm{\sigma}\big(\slogL[(i)]\big) \mleft(\grad\log \probD{\bm{\tilde\itemRel}^{(i)}|\betaParamsQD} + \grad\log \probD{\bm{\tilde\posBias}^{(i)}|\betaParamsPB} \mright).
	\nonumber
\end{align}

For the conditioned estimator (\ref{eq:method:condition}), we separately estimate the gradient for the position bias parameters ($\gradPB$) as for the parameters of the relevance model ($\gradQD$).
Since the outermost expectation in (\ref{eq:method:condition}) is over position bias samples, the $\gradPB$ gradient after self-normalization is quite straightforward:
\begin{equation}
\gradPB\log\big(\sLLD\big)  \!\approx\!  \frac{1}{S_\PB}  \!\sum_{i=1}^{S_\PB}\!  \Big(\bm{\sigma}\Big(\sum_\qd  \log \sJJqd[(i)]\Big)\gradPB\log \probD{\bm{\tilde\posBias}^{(i)}|\betaParamsPB}\!\Big)
.
\end{equation}
We first rewrite the gradient $\gradQD$ of (\ref{eq:method:condition}) without self-normalization:

\begin{equation}
	\begin{split}
	\gradQD\LLD
	&=\gradQD \mathop{\mathds{E}}_{\posBiasBM}\Big[\prod_\qd \JJqd\Big]
	=\gradQD \mathop{\mathds{E}}_{\posBiasBM}\Big[\exp\!\Big(\sum_\qd \log\big(\JJqd\big)\Big) \Big]
	 \\[-1ex]
	&= \mathop{\mathds{E}}_{\posBiasBM}\Big[\gradQD\exp\!\Big(\sum_\qd \log\big(\JJqd\big)\Big) \Big] \\[-1ex]
	&= \mathop{\mathds{E}}_{\posBiasBM}\Big[\exp\!\Big(\sum_\qd \log\big(\JJqd\big)\Big)\sum_{\qd} \gradQD \log\big(\JJqd\big) \Big]\\[-1ex]
	&= \mathop{\mathds{E}}_{\posBiasBM}\Big[\exp\!\Big(\sum_\qd \log\big(\JJqd\big)\Big)\sum_{\qd} \frac{\gradQD \JJqd}{\JJqd} \Big].
	\end{split}
	\label{eq: Grad QD - Product rewritten}
\end{equation}
This leads us to the following self-normalized gradient estimator:
\begin{equation}
		\gradQD \log\big(\sLLD\big) \approx \frac{1}{S_\PB}\sum_{i=1}^{S_\PB}\mleft(\bm{\sigma}\bigg(\sum_\qd \log\big(\sJJqd[(i)]\big)\bigg)\sum_{\qd}\frac{\gradQD\sJJqd[(i)]}{\sJJqd[(i)]}\mright)\,,
\end{equation}
where $\gradQD\sJJqd[(i)]$ is an estimate for the sample $\posBiasBM^{(i)}$:
\begin{equation}
	\gradQD\,\sJJqd[(i)] \approx \frac{1}{S_\QD}\!\sum_{j=1}^{S_\QD} \sJqd[(i,j)]\gradQD\log(\probD{\tilde\itemRel_\qd^{(j)}|\alpha_\qd,\beta_\qd}).
\end{equation}
\begin{figure*}[t]
	\centering
	{
		\renewcommand{\arraystretch}{0.5}
		\setlength\tabcolsep{0.01pt}
		\begin{tabular}{r c c r}
			& \vspace{-0.05mm} \footnotesize Istella-S &  \footnotesize MSLR-Web10k \\
			\raisebox{3.5\normalbaselineskip}[0pt][0pt]{\rotatebox[origin=c]{90}{
					\footnotesize Log-Likelihood
			}}
			&
			\includegraphics[scale=0.45]{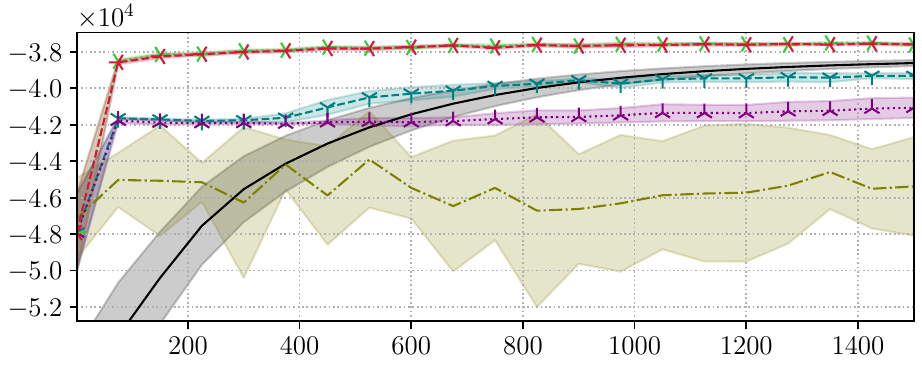}
			&
			\includegraphics[scale=0.45]{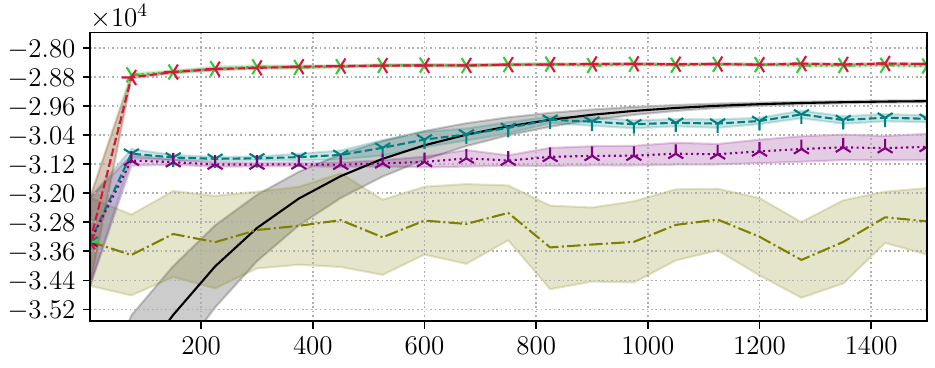}
			&
			\raisebox{1.25\normalbaselineskip}[0pt][0pt]{
				\includegraphics[scale=0.45]{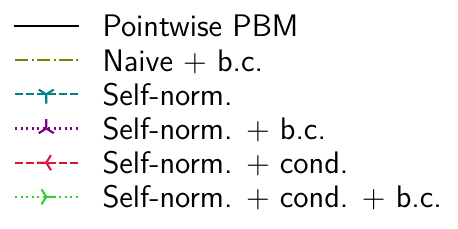}
			}
		\end{tabular}
	}
	\vspace{-0.6\baselineskip}
	\caption{Log-likelihood of test-set clicks as predicted by pointwise and epistemic \acp{PBM} trained with different methods over 1500 epochs. X-axis: training epochs; Y-axis: mean log-likelihood of clicks on test-set. The shaded areas indicate 95\% confidence intervals computed using a Student's $t-$distribution with 14 degrees of freedom. Results are based on 15 independent runs.}
	\label{fig: Plots Log-Likelihood}
\end{figure*}

Thus, through numerically-stable self-normalization, we deal with near-zero likelihoods for samples; equivalently, self-normalization results in the gradients of the log-likelihood.
Thereby, we increase the stability of both our standard and conditioned gradient estimates, which should result in more robust optimization.

\subsection{Entropy regularization}
Section~\ref{sec:background:entropy} discussed the common practice of entropy regularization for evidential deep learning.
Due to the specifics of our setup, we slightly adapt this approach before applying it.
First, we note that whilst there are many more query-item pairs than positions ($|\mathcal{D}| >> K$), the individual position-bias parameters $\posBias_k$ are involved in many more click probabilities than the $\itemRel_\qd$.
Nevertheless, standard entropy (\ref{eq:background:entropy}) gives equal weight to all parameters, we correct for this imbalance by introducing a modified entropy $\bar{\mathbf{H}}'$ (cf.\ (\ref{eq:background:entropybeta})):
\begin{equation}
	\bar{\mathbf{H}}'\!(\betaParamsQD,\betaParamsPB)=\frac{1}{\QDsize}\!\sum_{\qd}H(\alpha_\qd,\beta_\qd)+\frac{1}{K}\!\sum_k H(\alpha_k, \beta_k)
	.
\end{equation}
Furthermore, as our main computations are all in log-space, we add the regularization term to the log-likelihood, with $\lambda \in \mathds{R}_{\geq 0}$:
\begin{equation}
	\log\!\big(\tilde{\mathcal{L}}'\big)=\log\!\big(\tilde{\mathcal{L}}(\betaParamsQD, \betaParamsPB)\big)+\lambda \bar{\mathbf{H}}'\!(\betaParamsQD,\betaParamsPB)
	.
\end{equation}
Thereby, when $\lambda > 0$, our optimization aims to find parameters that balance maximizing the likelihood of the observed clicks and maximizing the epistemic uncertainty in the model.
Thus, if there are multiple values for the position bias ($\posBiasBM$) and relevance ($\itemRelBM$) variables that explain the data similarly well, then the entropy regularization pushes the optimization to spread epistemic probability over all of these values, avoiding a potential model collapse.

\section{Experimental Setup}
\label{sec:experimentalsetup}
We perform several experiments to evaluate how accurately our epistemic \ac{PBM} can predict clicks and whether its epistemic distributions capture uncertainty appropriately.
Specifically: we compare our proposed techniques: \emph{self-normalization}, \emph{conditioning on position bias} and \emph{entropy regularization} (Section~\ref{sec:method}) and a traditional pointwise \ac{PBM}.
We apply a \emph{semi-synthetic} setup because our evaluation requires a comparison between learned click models and ground-truth parameter values.
For reproducibility, our implementation will be made available upon publication.

\noindent\textbf{Datasets.}\; We perform our experiments on two learning-to-rank datasets: \emph{MSLR-Web-10K}~\citep{MSLR_WEB} and the \emph{Istella-S}~\citep{Istella}.
MSLR contains 30,000 queries with an average of 125 pre-selected documents per query and 136 features to represent query-document pairs, we only use fold-1 of the dataset.
Istella has 33,018 queries, an average of 103 pre-selected documents per query and 220 features.
Both dataset have an expert-judged relevance label for every query-document pair: $y_\qd \in \{0,1,2,3,4\}$ where $0$ is the least relevant and $4$ the most.
The datasets come with standard training/validation/test set splits.

\noindent\textbf{Click generation.}\;
For each dataset, we simulate $10^5$ query-inter\-actions by uniformly sampling queries over the train, validation and test sets (with replacement).
Simulated clicks are sampled from a \ac{PBM} with $\theta_k=\frac{1}{k}$ and  $\itemRel_\qd=0.9(y_\qd/4)+0.1$ on sampled top-5 rankings ($K=5$).
Our logging policy ranker is a neural-network (2 hidden layers of 64 units) Plackett-Luce ranking model~\citep{oosterhuis2021models} optimized on $30$ random training-set queries in a supervised manner. 

\noindent\textbf{Model parameters.}\;
Our click models are based on neural networks with 3 hidden layers of 64, 64 and 32 units to predict document relevances and a lookup table to predict position bias.
The epistemic neural networks output two parameters: the mean $\mu \in (0,1)$ (using a sigmoid activation) and confidence $\nu \in [1,\infty)$ (using the $|x| + 1$ transformation).
The predicted beta distributions parameters were then obtained by: $\alpha = \mu\nu$ and $\beta = (1-\mu)\nu$.
The pointwise \ac{PBM} baseline uses the same network structure but outputs the $\zeta \in (0,1)$ parameter directly.
All models are optimized over 1500 training epochs; we do not apply entropy regularization except when stated otherwise. For gradient estimation, we take $S=60$ samples when not applying conditioning, and $S_\PB=60$ and $S_\QD=60$ when conditioning (see Section~\ref{sec:method:conditioning}). Pointwise \ac{PBM} has an exact gradient.
During training, we batch the data into chunks of 12k and 35k $q$,$d$ pairs for epistemic and pointwise models respectively. 

\noindent\textbf{Evaluation.}\;
To evaluate the predictive accuracy of our optimized \ac{PBM}, we evaluate the log-likelihood of the clicks generated on the test-set.
We also want to evaluate whether the uncertainty captured by our epistemic models is appropriate, however, there is no standard method for evaluating epistemic uncertainty~\citep{hullermeier2021aleatoric, pmlr-v235-juergens24a}.
Thus, we plot the learned distributions for the $K=5$ position bias parameters ($\theta_k$) for visual inspection.
Due to the large number of query-document pairs, we cannot do the same for $\zeta_{q,d}$; instead we analyze the means and standard deviations of the relevance distributions.
All results are based on 15 independent runs.

\begin{figure*}[t]
	{
		\centering
		\renewcommand{\arraystretch}{0.001}
		\setlength\tabcolsep{0.01pt}
		\begin{tabular}{r r c c c c c}
			& & $\posBias_{1}$ & $\posBias_{2}$ & $\posBias_{3}$ & $\posBias_{4}$ & $\posBias_{5}$
			\\
    		\cline{1-2}
			\multirow{4}{*}{\raisebox{-2.25\normalbaselineskip}[0pt][0pt]{\rotatebox[origin=c]{90}{\footnotesize Istella-S}}}
			& \raisebox{2.2\normalbaselineskip}[0pt][0pt]{\rotatebox[origin=c]{90}{
					\footnotesize Na\"ive+bc.
			}} & \includegraphics[scale=0.3195]{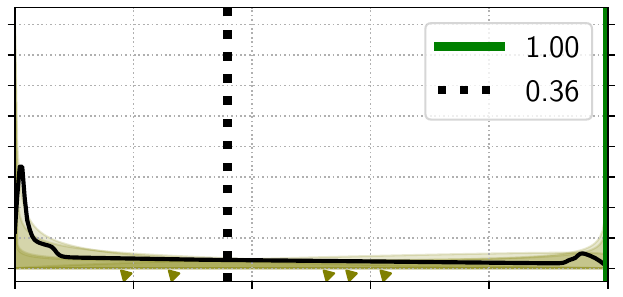} 
			& \includegraphics[scale=0.3195]{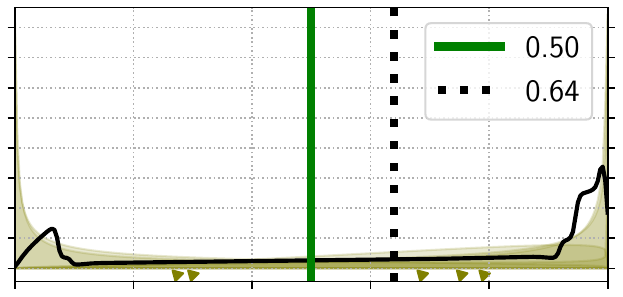} 
			& \includegraphics[scale=0.3195]{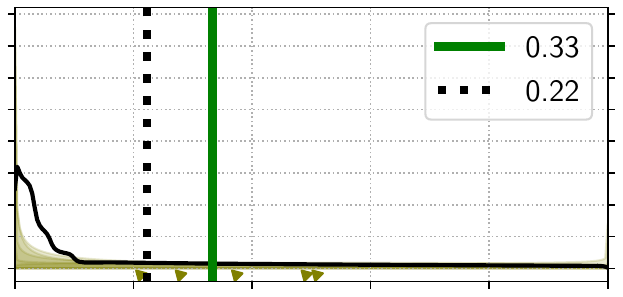} 
			& \includegraphics[scale=0.3195]{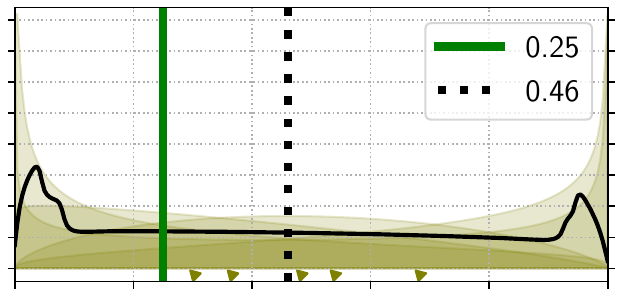}
			& \includegraphics[scale=0.3195]{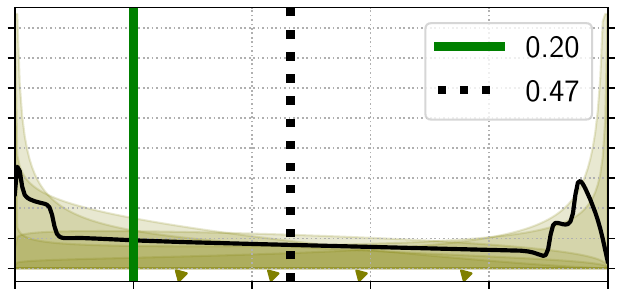}
			\\
			& \raisebox{2.25\normalbaselineskip}[0pt][0pt]{\rotatebox[origin=c]{90}{
					\footnotesize Self-norm.
			}}  & \includegraphics[scale=0.3195]{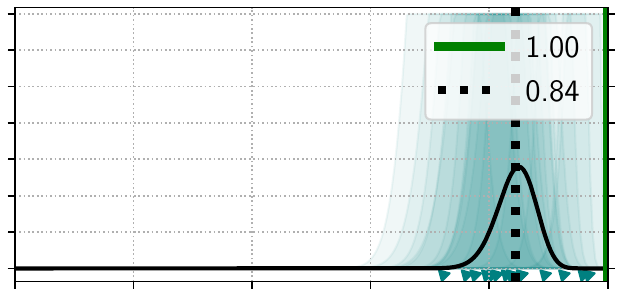} 
			& \includegraphics[scale=0.3195]{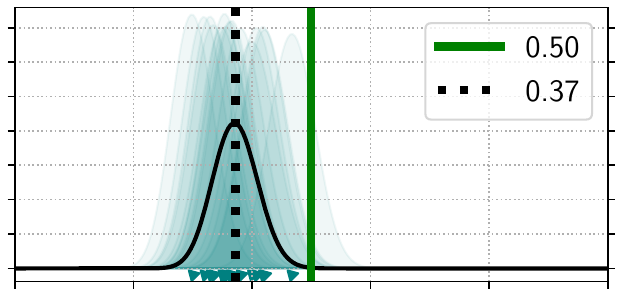} 
			& \includegraphics[scale=0.3195]{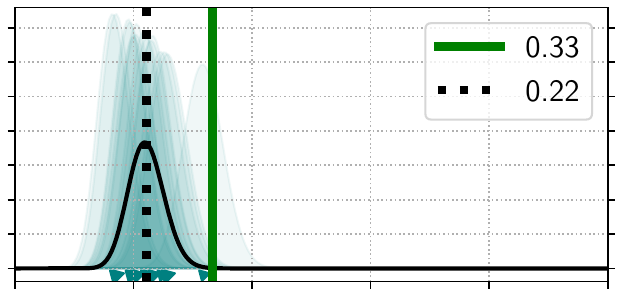} 
			& \includegraphics[scale=0.3195]{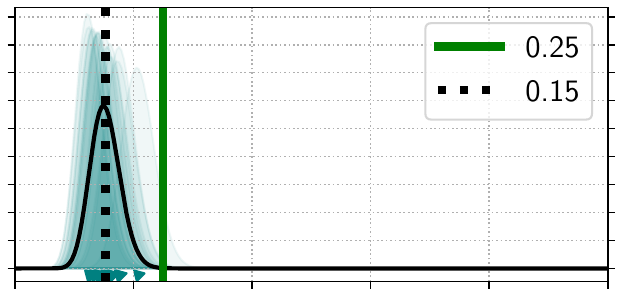}
			& \includegraphics[scale=0.3195]{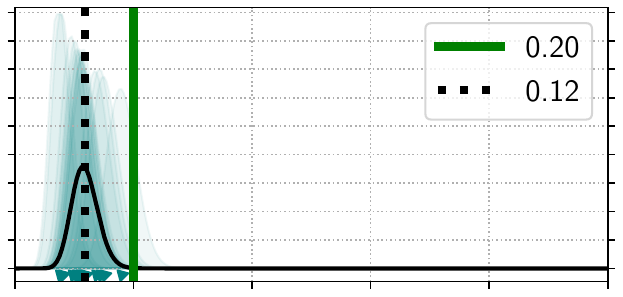}
			\\
			& \raisebox{2\normalbaselineskip}[0pt][0pt]{\rotatebox[origin=c]{90}{
					\footnotesize S.-norm.+cond.
			}}  & \includegraphics[scale=0.3195]{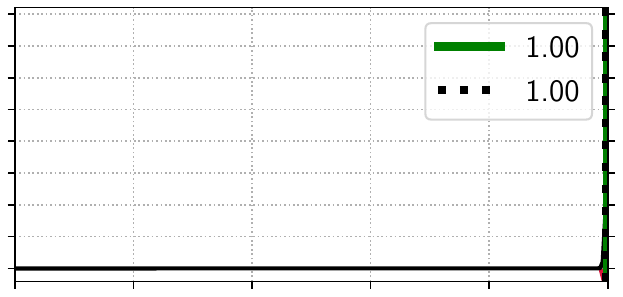} 
			& \includegraphics[scale=0.3195]{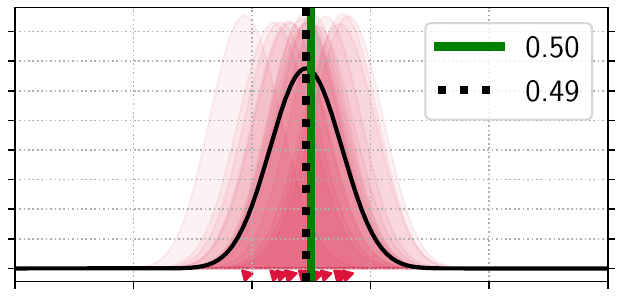} 
			& \includegraphics[scale=0.3195]{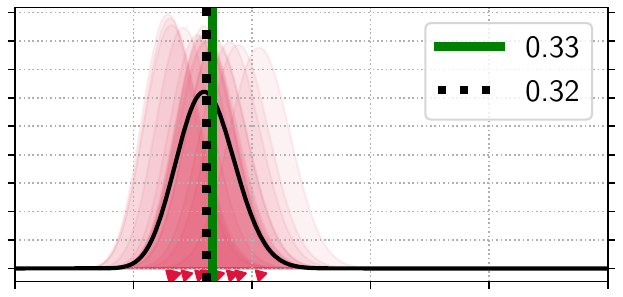} 
			& \includegraphics[scale=0.3195]{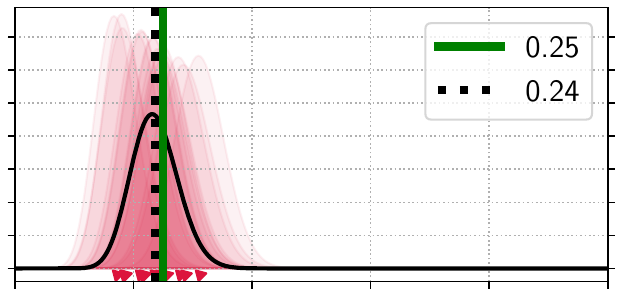}
			& \includegraphics[scale=0.3195]{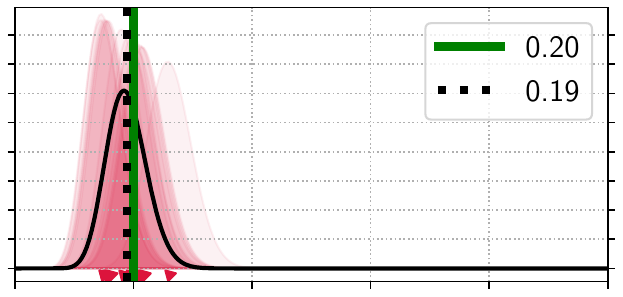}
			\\
			& \footnotesize \raisebox{0.8\normalbaselineskip}[0pt][0pt]{p.w.}
			& \includegraphics[scale=0.319]{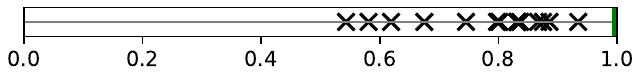} 
			& \includegraphics[scale=0.319]{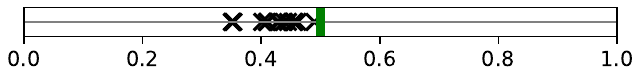} 
			& \includegraphics[scale=0.319]{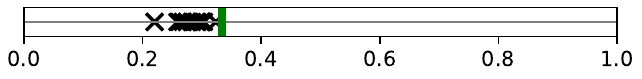} 
			& \includegraphics[scale=0.319]{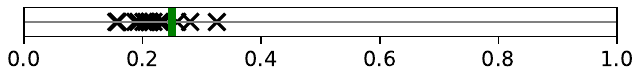}
			& \includegraphics[scale=0.319]{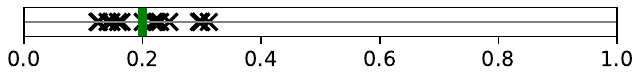}	
			\\
			\cline{1-2}
			\multirow{4}{*}{\raisebox{-2.05\normalbaselineskip}[0pt][0pt]{\rotatebox[origin=c]{90}{\footnotesize MSLR-Web10k}}}
			& \raisebox{2.2\normalbaselineskip}[0pt][0pt]{\rotatebox[origin=c]{90}{
					\footnotesize Na\"ive+bc.
			}} & \includegraphics[scale=0.3195]{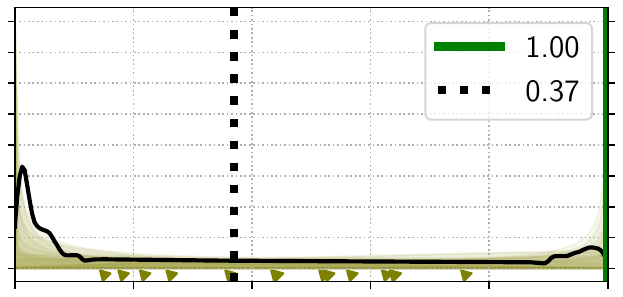} 
			& \includegraphics[scale=0.3195]{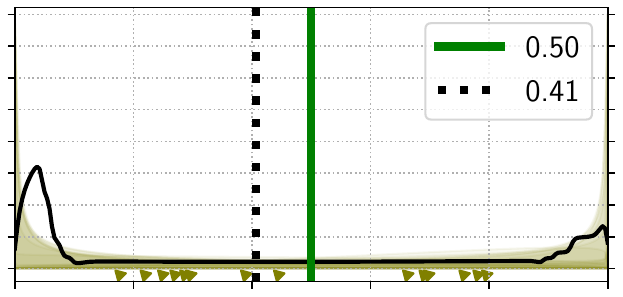} 
			& \includegraphics[scale=0.3195]{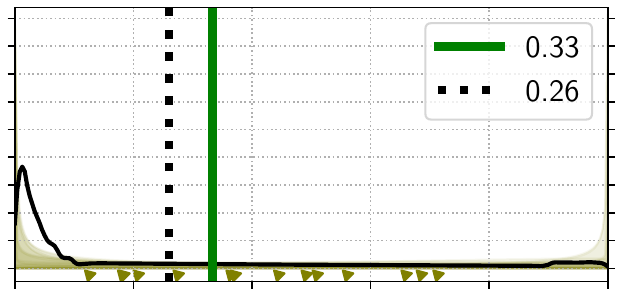} 
			& \includegraphics[scale=0.3195]{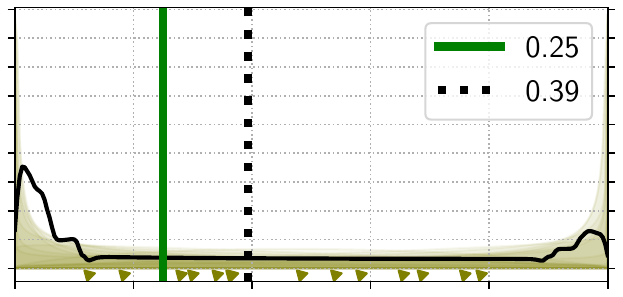}
			& \includegraphics[scale=0.3195]{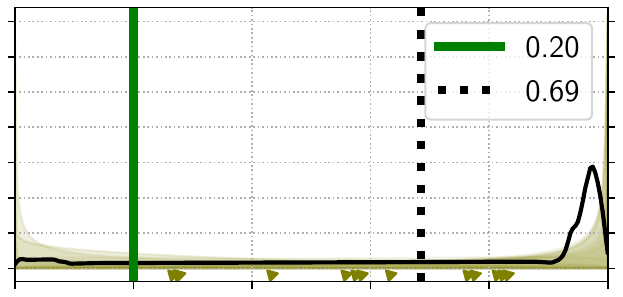}
			\\
			& \raisebox{2.25\normalbaselineskip}[0pt][0pt]{\rotatebox[origin=c]{90}{
					\footnotesize Self-norm.
			}}  & \includegraphics[scale=0.3195]{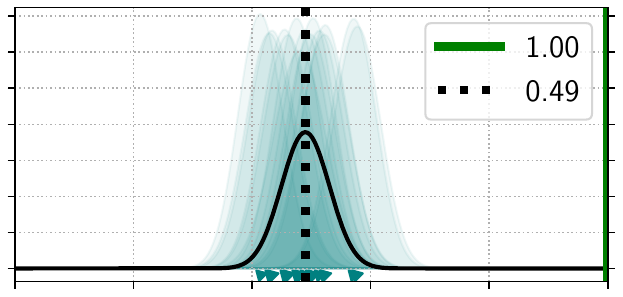} 
			& \includegraphics[scale=0.3195]{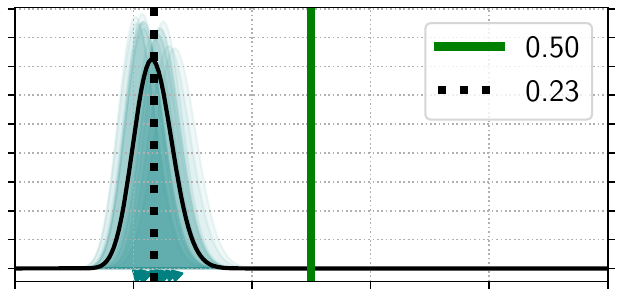} 
			& \includegraphics[scale=0.3195]{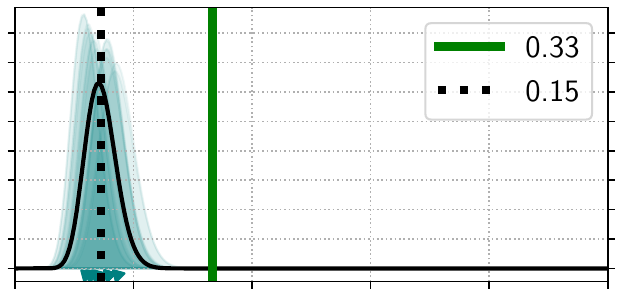} 
			& \includegraphics[scale=0.3195]{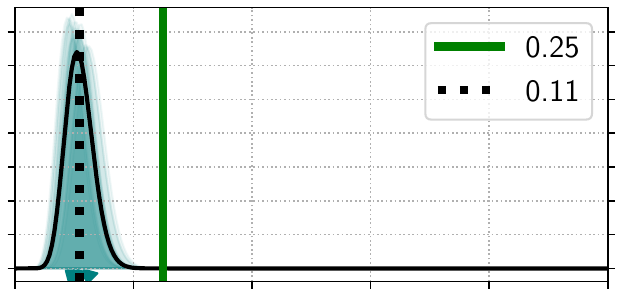}
			& \includegraphics[scale=0.3195]{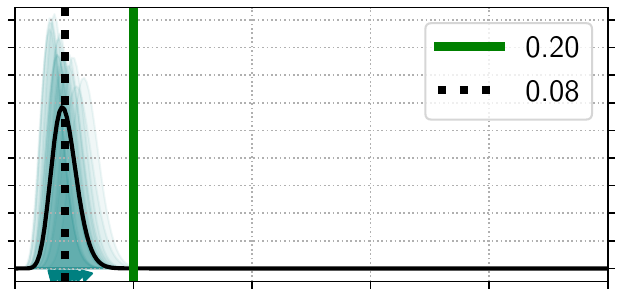}
			\\
			& \raisebox{2\normalbaselineskip}[0pt][0pt]{\rotatebox[origin=c]{90}{
					\footnotesize S.-norm.+cond.
			}}  & \includegraphics[scale=0.3195]{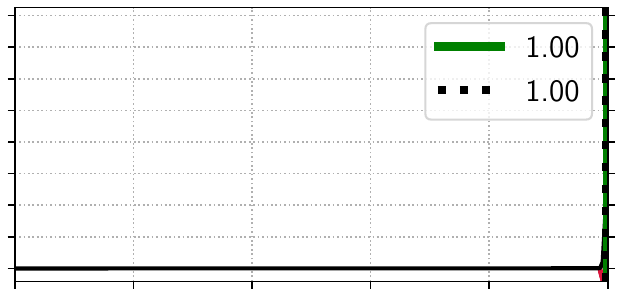} 
			& \includegraphics[scale=0.3195]{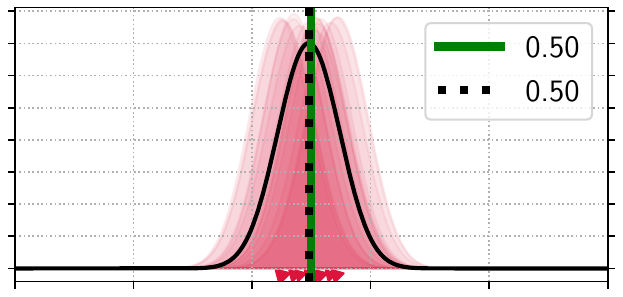} 
			& \includegraphics[scale=0.3195]{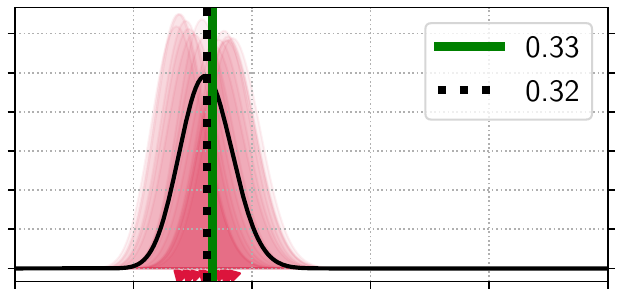} 
			& \includegraphics[scale=0.3195]{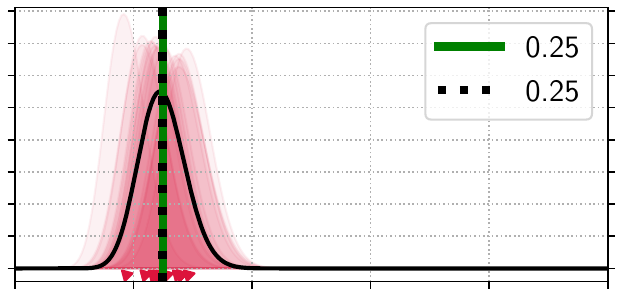}
			& \includegraphics[scale=0.3195]{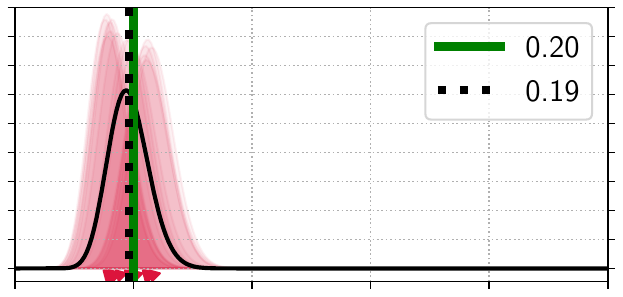}
			\\
			& \footnotesize \raisebox{0.8\normalbaselineskip}[0pt][0pt]{p.w.}
			& \includegraphics[scale=0.319]{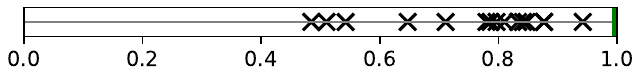} 
			& \includegraphics[scale=0.319]{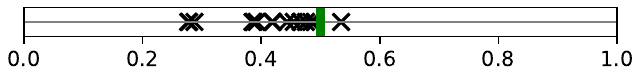} 
			& \includegraphics[scale=0.319]{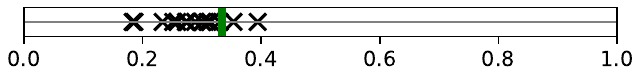} 
			& \includegraphics[scale=0.319]{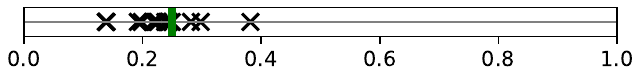}
			& \includegraphics[scale=0.319]{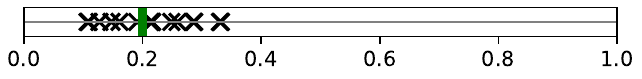}
			\\
			\cline{1-2}
			\multicolumn{7}{c}{\includegraphics[scale=0.4]{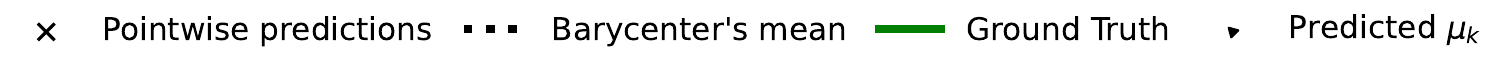}}
		\end{tabular}
		\vspace{-0.7\baselineskip}
		\caption{Predicted distributions by epistemic \acp{PBM} optimized with different methods and pointwise \ac{PBM} predictions.} 
		\label{fig: Position Bias Distributions MSLRWEB10k}
		\vspace{-1.0\baselineskip}
	}
\end{figure*}

\section{Experimental Results}

\subsection{Comparison of proposed techniques}
\label{sec:results:comparison}
The main purpose of click models is making accurate predictions; thus our first research question is: \emph{Which of our proposed techniques result in a higher log-likelihood of our epistemic click model?}

Figure~\ref{fig: Plots Log-Likelihood} displays learning curves in terms of log-likelihood over training epochs of models optimized with different gradient estimation techniques.
We see that the general trends and differences between methods is very consistent between the Istella and MSLR datasets:
The na\"ive estimator struggles to improve log-likelihood much beyond its starting point, suggesting that its gradient estimation is too unstable to be useful for optimization (even with baseline-corrections applied).
Self-normalization results in a higher log-likelihood, and surprisingly, baseline-corrections degrades its point of convergence.
Finally, self-normalization and conditioning on position bias leads to the highest log-likelihood, with or without baseline-corrections.
The differences between methods are substantial and the 95\% confidence intervals indicate that they are statistically significant.
Overall, these results strongly suggest that the stability of the gradient is a serious issue for learning epistemic click models and that our techniques effectively increase stability.

Therefore, we have a clear answer to our first research question:
\emph{Self-normalization and conditioning on position bias lead to a considerably higher log-likelihood than na\"ive estimation;  baseline corrections have no positive effect when applied with self-normalization.}

\subsection{Comparison with pointwise PBM}
Next, we consider whether there is a trade-off between modeling uncertainty and predictive accuracy.
Accordingly, our second research question is:
\emph{Can our epistemic click model reach the same log-likelihood as a traditional pointwise click model?}

We again turn to Figure~\ref{fig: Plots Log-Likelihood} where we can see the log-likelihood curves of the pointwise \ac{PBM} and the epistemic \ac{PBM} with self-normalization and conditioning.
Surprisingly, we see that the epistemic click model converges at a higher log-likelihood and requires far fewer training epochs to reach convergence;
The epistemic \ac{PBM} needs 100 epochs, whereas the pointwise \ac{PBM} does not even converge within 1200 epochs.
Thus,  instead of a trade-off, these results show that it actually improves accuracy.
We speculate that this improvement comes from differences in initialization:
Pointwise \acp{PBM} are initialized with specific position bias parameters or models that gives pointwise relevance predictions; in our experiments, default logit-initialization results in initial values around 0.5 for all parameters.
Conversely, the epistemic \ac{PBM} can be initialized with an almost-uniform distribution over the possible parameter values; in our experiments, this happens because $\nu$ values start around 1.
As a result, our epistemic \ac{PBM} starts from a neutral initial position, and furthermore, by predicting distributions it can keep multiple possible parameter values in consideration.
Possibly, this could also allow it to be less prone to get stuck in local optima than the pointwise \ac{PBM}.
This matches \citep[Chapter 4.2]{chuklin2015basic} which stresses the effect of intialization on the convergence of pointwise click models.

Regardless of the underlying reason, we answer the second research question in the affirmative:
\emph{Epistemic click models can provide higher log-likelihood than traditional pointwise click models.}

\begin{figure*}[t]
	\centering
	{
		\renewcommand{\arraystretch}{0.001}
		\setlength\tabcolsep{0.01pt}
		\begin{tabular}{r c c c c}
			& \hspace{5mm}\footnotesize Average of the $\mu_\qd$
			& \hspace{5mm}\footnotesize s.s.d of the $\mu_\qd$
			& \hspace{5mm}\footnotesize Average of the $\nu_\qd$
			& \hspace{5mm}\footnotesize s.s.d of the $\nu_\qd$
			\\
			\raisebox{2.65\normalbaselineskip}[0pt][0pt]{\rotatebox[origin=c]{90}{\footnotesize Istella-S}}
			&\includegraphics[scale=0.45]{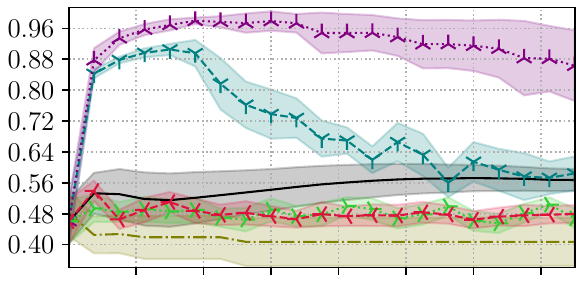}
			& \includegraphics[scale=0.45]{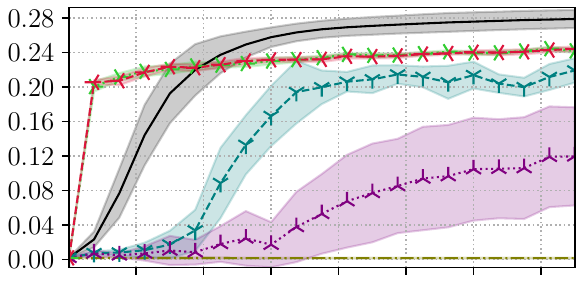}
			& \includegraphics[scale=0.45]{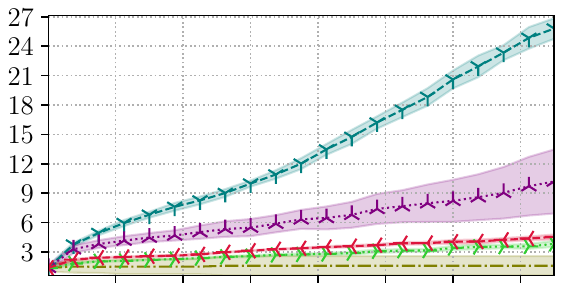}
			& \includegraphics[scale=0.45]{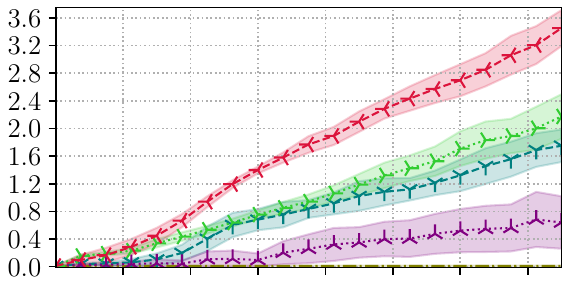}\\
			\raisebox{3.35\normalbaselineskip}[0pt][0pt]{\rotatebox[origin=c]{90}{\footnotesize MSLR-Web10K}}
			&\includegraphics[scale=0.45]{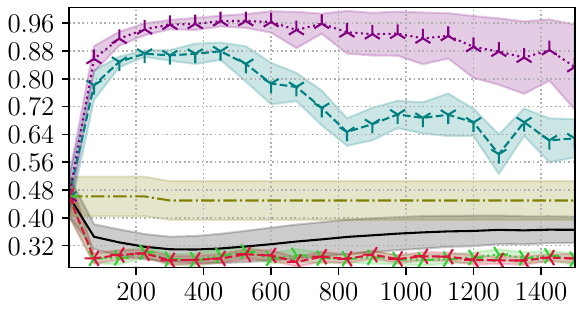}
			& \includegraphics[scale=0.45]{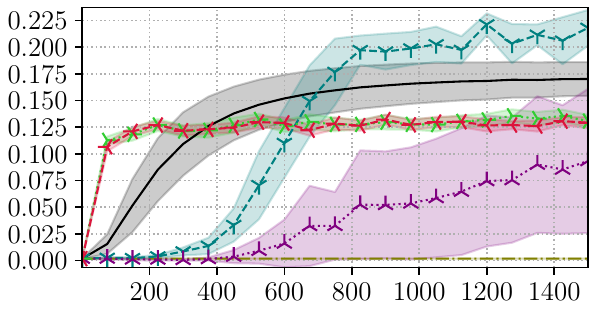}
			& \includegraphics[scale=0.45]{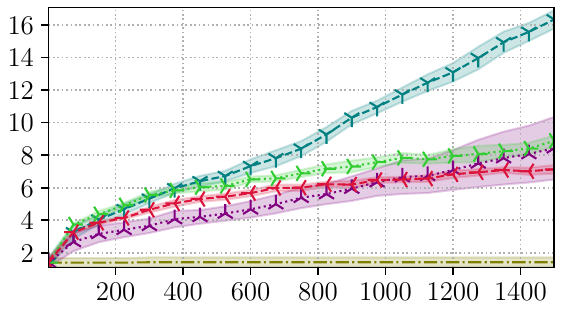}
			& \includegraphics[scale=0.45]{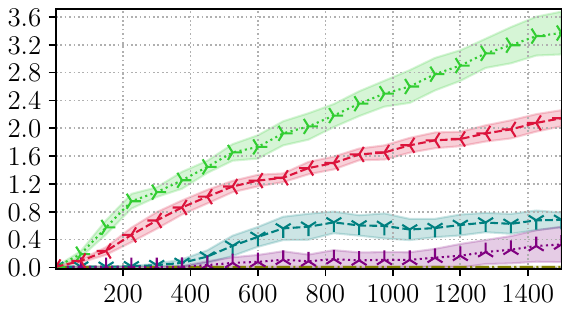}
			\\
			\multicolumn{5}{c}{
			\includegraphics[scale=0.44]{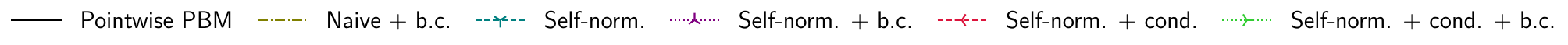}
			}
		\end{tabular}	

	}
	\vspace{-0.6\baselineskip}
	\caption{Statistics of the relevance distributions of the epistemic \ac{PBM} during optimization. X-axis: training epoch; Y-axis: average or sampled standard deviation of the predicted means $\mu_\qd$ and confidences $\nu_\qd$ over all the query-document pairs $\qd$.}
	\label{fig: Plots Relevance parameters}
	\vspace{-1\baselineskip}
\end{figure*}

\subsection{The learned epistemic distributions.}

We continue by asking the third research question: \emph{Does our epistemic \ac{PBM} appropriately capture epistemic uncertainty?}

Figure~\ref{fig: Position Bias Distributions MSLRWEB10k} displays the predicted distributions for the position bias parameters $\theta_k$ for epistemic \acp{PBM} optimized with several methods and the predictions of the pointwise \acp{PBM}.
For interpretability, a Barycenter over the distributions of the 15 runs is also visualized.

We see very clearly that na\"ive optimization does not lead to informative distributions, but extremely flat predicted distributions for all parameters.
Self-normalization \emph{without} conditioning results in more concentrated distributions but they systematically underestimate all parameter values.
In contrast, self-normalization \emph{with} conditioning on position bias leads to distributions concentrated around the ground truth values.
Finally, the pointwise estimates systematically underestimate position bias on the first three positions, but are more evenly spread around the correct values on the final two positions.
The location of the means of the distributions corresponds to their predictive accuracy, i.e., the methods that result in a higher likelihood  (as discussed in Section~\ref{sec:results:comparison}) also provide distributions more centered around the true position bias.

Additionally, Figure~\ref{fig: Position Bias Distributions MSLRWEB10k} shows that there is a noticeable spread in the values on which the methods converge; for the epistemic \acp{PBM} this seems less of an issue as their distributions still mostly overlap, but this is not applicable to the predictions of the pointwise \acp{PBM}.
Evaluating the width of the epistemic distributions is not straightforward, however, it is clear that na\"ive optimization results in distributions that are far too wide, conversely, self-normalization \emph{without} conditioning results in distributions that are too concentrated as the true values get almost no epistemic probability.
In contrast, the distributions self-normalization \emph{with} conditioning are placed around the true values, so give much probability to them, yet, it is unclear how to evaluate their width.
Because we know this model has the highest likelihood, it appears that its uncertainty around its position bias parameters provide better estimates than the other epistemic and pointwise \acp{PBM}.
It also appears to match the spread from the different pointwise \ac{PBM} estimates.
Thus, at most, we can conclude that \emph{our epistemic \ac{PBM} with self-normalization and conditioning appropriately captures uncertainty in its position bias predictions, as far as can be judged from visual inspection}.

The widths of the epistemic distributions over the relevance parameters are more difficult to analyze, as due to the enormous number of query-document pairs, we cannot visualize them all.
Instead, Figure~\ref{fig: Plots Relevance parameters} shows the average and the s.s.d.\ of the distributions' mean and confidence parameters $\mu_{q,d}$ and $\nu_{q,d}$ (see Section~\ref{sec:experimentalsetup}).
We see that the average of $\mu_{q,d}$ increases substantially during the first training epochs when \emph{not} conditioning the estimators on position bias, which seem incorrect as high relevance labels are rare in the datasets.
Moreover, we see that the $\mu_{q,d}$ s.s.d.\ for the na\"ive estimator is near zero, indicating that the same $\mu_{q,d}$ is used for all documents; clearly it has failed to learn meaningful parameters.
When we consider $\nu_{q,d}$ we see that the na\"ive model has extremely low confidence on average, and a near-zero s.s.d.; again indicating that all documents get the same confidence.
The self-normalized estimator \emph{without} baseline corrections continues to increase its average $\nu_{q,d}$, resulting in extremely high confidence levels.
Self-normalization \emph{with} conditioning leads to high s.s.d.\ for both $\mu_{q,d}$ and $\nu_{q,d}$, indicating it varies the means and confidence parameters over different query-document pairs; the average values appear to take reasonable values for the datasets.
Thus, similar to our previous conclusions, we conclude that \emph{self-normalization \emph{with} conditioning results in plausible relevance distributions of our epistemic \ac{PBM}, whereas without these techniques this is not the case.}

Lastly, we answer our third research questions as follows: \emph{With self-normalization and conditioning, the epistemic \ac{PBM} appear to capture appropriate levels of epistemic uncertainty in its predictions over both the position bias and relevance parameters.}

\begin{figure*}[t]
	{
		\centering
		\renewcommand{\arraystretch}{0.001}
		\setlength\tabcolsep{0.01pt}
		\begin{tabular}{r c c c c c}
			& \hspace{8mm}\footnotesize Log-Likelihood
			& \hspace{7mm}\footnotesize\footnotesize Average of $\mu_\qd$
			& \hspace{7mm}\footnotesize\footnotesize s.s.d. of $\mu_\qd$
			& \hspace{7mm}\footnotesize\footnotesize Average of $\nu_\qd$
			& \hspace{7mm}\footnotesize\footnotesize s.s.d. of $\nu_\qd$
			\\
			\raisebox{2.5\normalbaselineskip}[0pt][0pt]{\rotatebox[origin=c]{90}{
					\footnotesize Istella-S
			}} 
			& \includegraphics[scale=0.45]{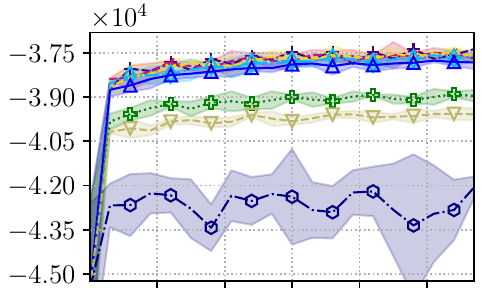}
			& \includegraphics[scale=0.45]{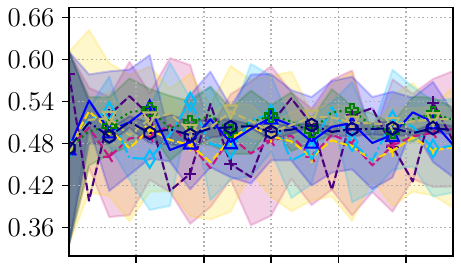} 
			& \includegraphics[scale=0.45]{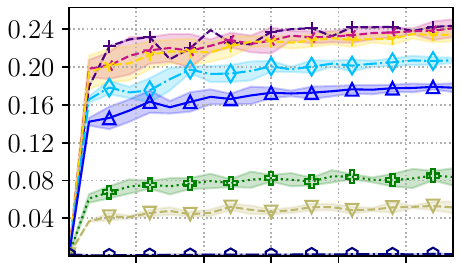} 
			& \includegraphics[scale=0.45]{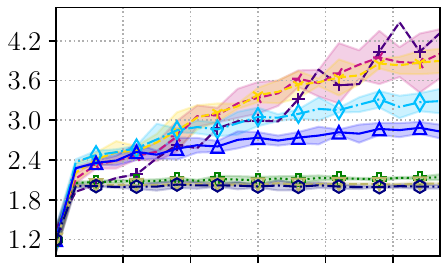} 
			& \includegraphics[scale=0.45]{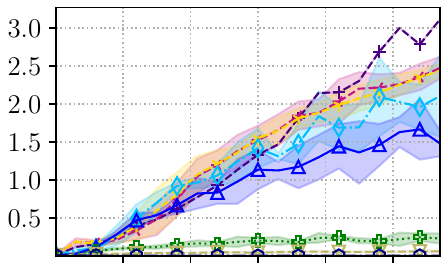}
			\\
			\raisebox{3.25\normalbaselineskip}[0pt][0pt]{\rotatebox[origin=c]{90}{
					\footnotesize MSLR-Web10k
			}} 
			& \includegraphics[scale=0.45]{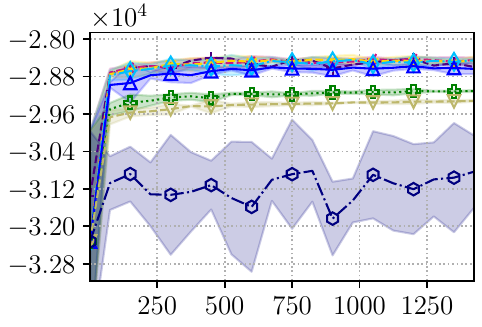}
			& \includegraphics[scale=0.45]{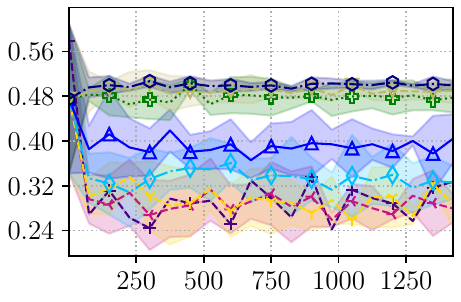} 
			& \includegraphics[scale=0.45]{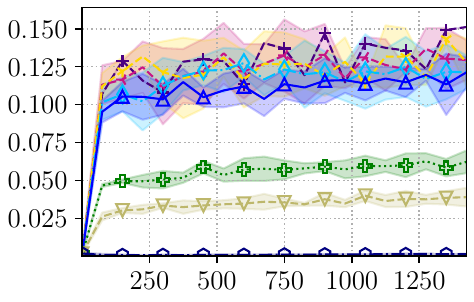} 
			& \includegraphics[scale=0.45]{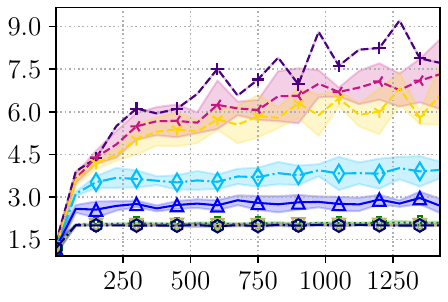} 
			& \includegraphics[scale=0.45]{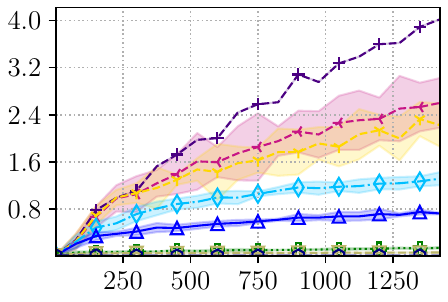}
			\\
			\multicolumn{6}{c}{
				\includegraphics[scale=0.42]{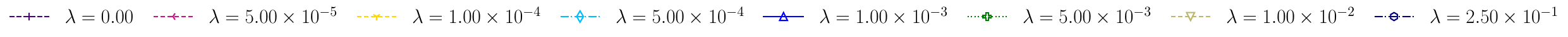}
			}
		\end{tabular}
		\vspace{-0.7\baselineskip}
		\caption{The effect of regularization on various statistics. X-axis: training epoch; Y-Axis: values of the indicated metrics.}
		\label{fig: Regularization impact}
		\vspace{-0.5\baselineskip}
	}
\end{figure*}
\begin{figure*}[t]
	\centering
	\renewcommand{\arraystretch}{0.5}
	\setlength\tabcolsep{0.001pt}
	\begin{tabular}{r r c c c c c}
		& & $\posBias_{1}$ & $\posBias_{2}$ & $\posBias_{3}$ & $\posBias_{4}$ & $\posBias_{5}$
		\\
		\multirow{2}{*}{\raisebox{0.2\normalbaselineskip}[0pt][0pt]{\rotatebox[origin=c]{90}{\footnotesize Istella-S}}}\,
		& \raisebox{2.25\normalbaselineskip}[0pt][0pt]{\rotatebox[origin=c]{90}{
				\footnotesize $\lambda=0.001$
		}}  & \includegraphics[scale=0.321]{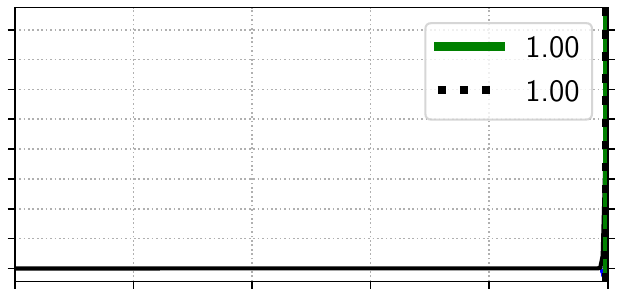} 
		& \includegraphics[scale=0.321]{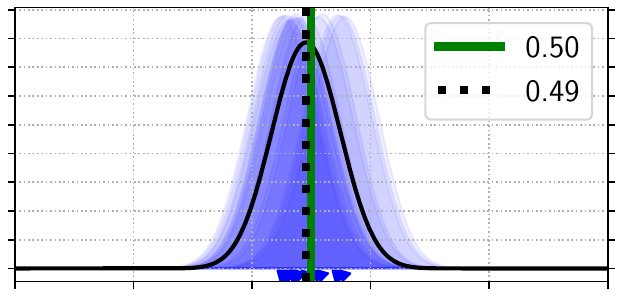} 
		& \includegraphics[scale=0.321]{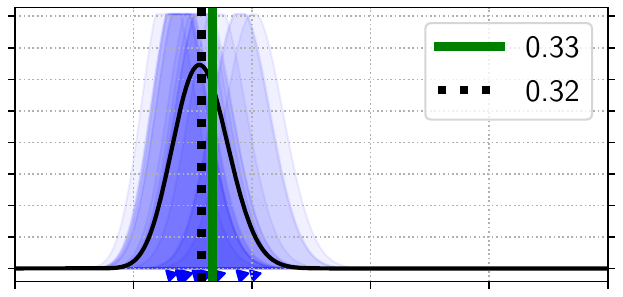} 
		& \includegraphics[scale=0.321]{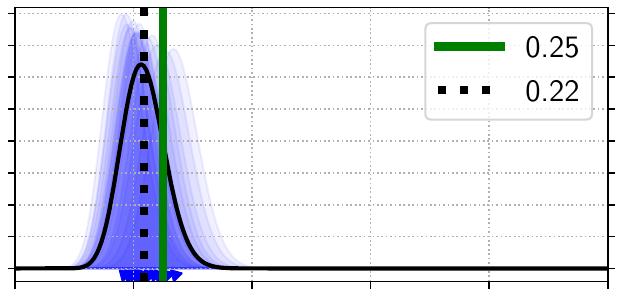}
		& \includegraphics[scale=0.321]{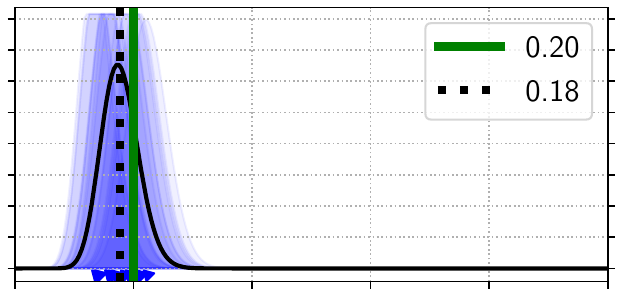}
		\\
		& \raisebox{2.5\normalbaselineskip}[0pt][0pt]{\rotatebox[origin=c]{90}{
				\footnotesize $\lambda=0.25$
		}}  & \includegraphics[scale=0.321]{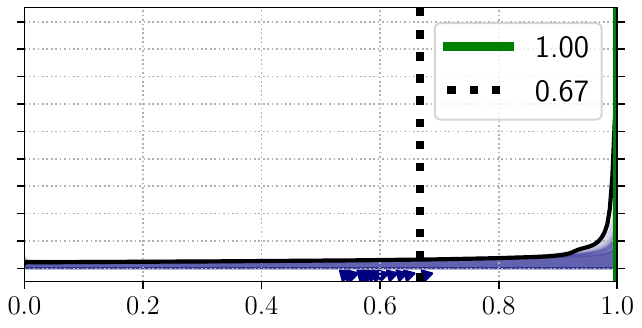} 
		& \includegraphics[scale=0.321]{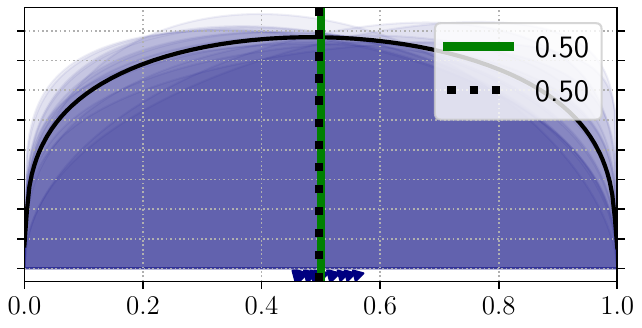} 
		& \includegraphics[scale=0.321]{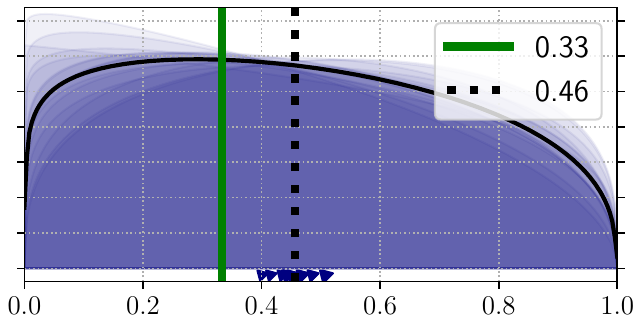} 
		& \includegraphics[scale=0.321]{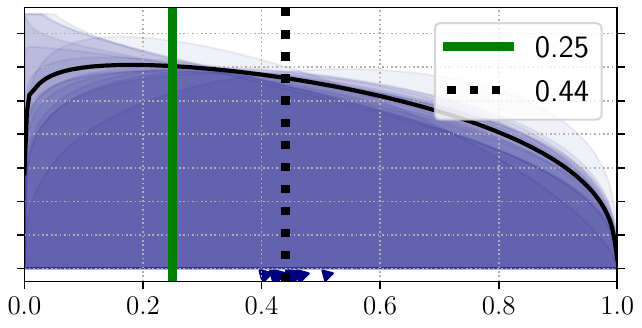}
		& \includegraphics[scale=0.321]{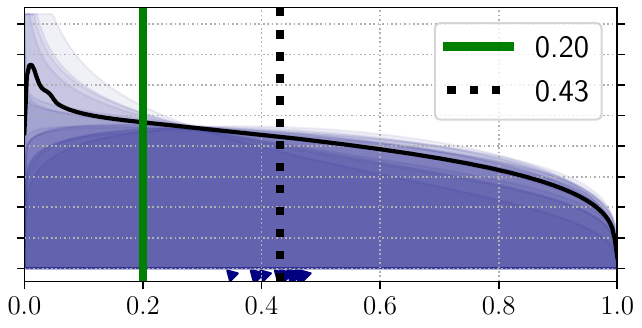}
		\\
		\multirow{2}{*}{\raisebox{0.2\normalbaselineskip}[0pt][0pt]{\rotatebox[origin=c]{90}{\footnotesize MSLR-Web10k}}}\,
		& \raisebox{2.15\normalbaselineskip}[0pt][0pt]{\rotatebox[origin=c]{90}{
				\footnotesize $\lambda=0.001$
		}}  & \includegraphics[scale=0.321]{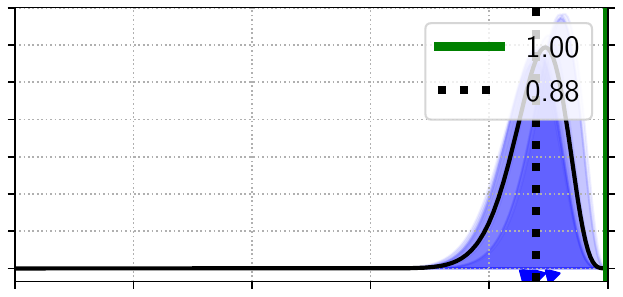} 
		& \includegraphics[scale=0.321]{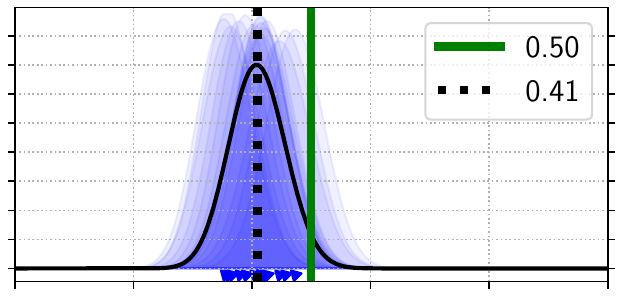} 
		& \includegraphics[scale=0.321]{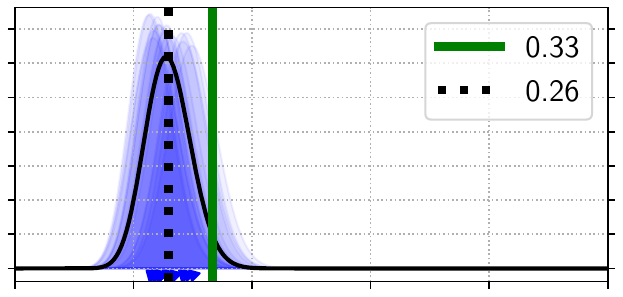} 
		& \includegraphics[scale=0.321]{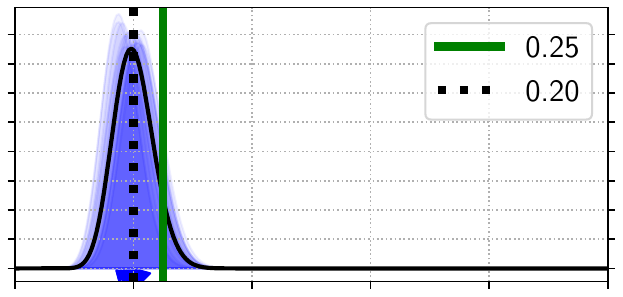}
		& \includegraphics[scale=0.321]{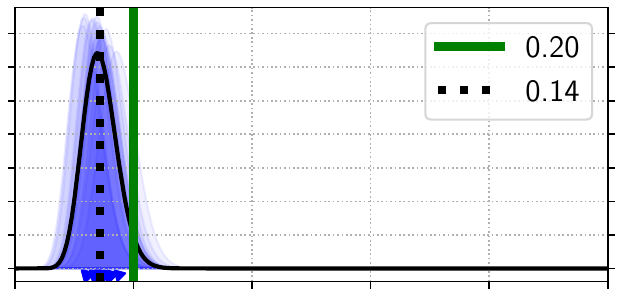}\\
		& \raisebox{2.5\normalbaselineskip}[0pt][0pt]{\rotatebox[origin=c]{90}{
				\footnotesize $\lambda=0.25$
		}}  & \includegraphics[scale=0.321]{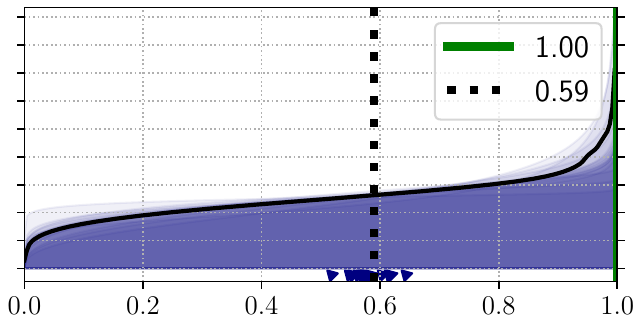} 
		& \includegraphics[scale=0.321]{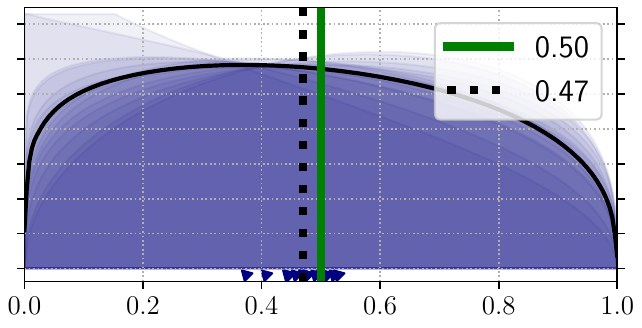} 
		& \includegraphics[scale=0.321]{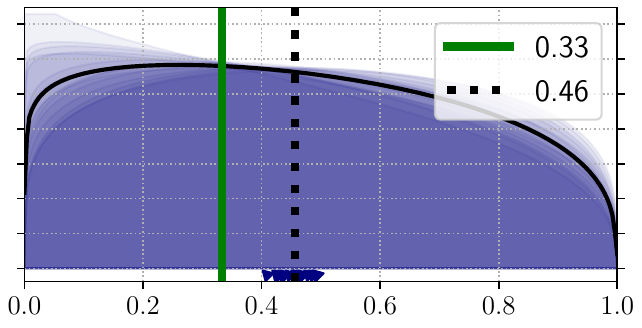} 
		& \includegraphics[scale=0.321]{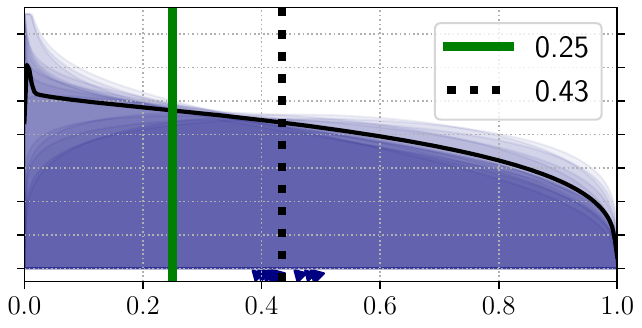}
		& \includegraphics[scale=0.321]{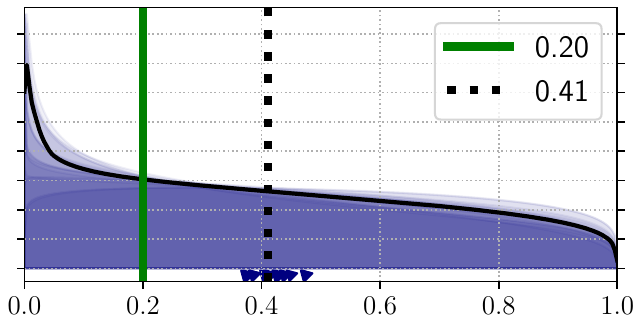}
	\end{tabular}
	\vspace{-0.7\baselineskip}
	\caption{Position bias distributions obtained with entropy regularization applied with weight $\lambda=0.001$ and $\lambda=0.25$ (cf.\ Figure~\ref{fig: Position Bias Distributions MSLRWEB10k}). }
	\label{fig: Position Bias REG Distributions MSLRWEB10k}
	\vspace{-1.15\baselineskip}
\end{figure*}

\subsection{The effect of entropic regularization}
So far we have not applied entropic regularization and only optimized log-likelihood.
To address this gap, we ask our fourth research question: \emph{How does entropic regularization impact epistemic \acp{PBM}?}

Figure~\ref{fig: Regularization impact} displays various statistics of epistemic \acp{PBM} trained with self-normalization and conditioning under different levels of entropic regularization.
Furthermore, Figure~\ref{fig: Position Bias REG Distributions MSLRWEB10k} displays the position bias distributions under mild and extreme entropy regularization.
From Figure~\ref{fig: Regularization impact}, we see that $\lambda \leq 10^{-3}$ does not lead to meaningful decreases in log-likelihood, but does result in lower average $\nu_{q,d}$.
Therefore, it appears that the models can have less confidence in their relevance predictions, while still maintaining predictive accuracy.
Figure~\ref{fig: Position Bias REG Distributions MSLRWEB10k} shows us that $\lambda = 10^{-3}$ affects the means of the position bias distributions, and some distributions are wider (i.e., $\theta_1$ on MSLR) but most appear very similar to when $\lambda=0$ (3rd and 7th rows in Figure~\ref{fig: Position Bias Distributions MSLRWEB10k}).
Thus, it seems the means of the position bias distributions are shifted to account for the lower confidence in the relevance distributions.
If we consider more extreme regularization with $\lambda=0.25$, we see in Figure~\ref{fig: Regularization impact} that this considerably reduces the likelihood and average $\nu_{q,d,}$; correspondingly, Figure~\ref{fig: Position Bias REG Distributions MSLRWEB10k} shows us that $\lambda=0.25$ leads to uninformatively wide position bias distributions.

Based on these observations, we answer the fourth research question:
\emph{Entropy regularization results in epistemic \acp{PBM} with less confidence in their predictions, if applied with moderation the resulting models are able to maintain near-optimal likelihood, but extreme regularization leads to unusably unconfident models.}

\subsection{Downstream task: Ranker evaluation}

Finally, our fifth research question considers the usefulness of our epistemic approach for downstream tasks:
\emph{Can our epistemic \ac{PBM} provide appropriate uncertainty distributions for ranker evaluation?}

Our task is to predict the expected number of clicks for different rankers than the one used for the observed click data, thus involving a distribution shift.
As rankers, we select three individual features from each dataset: the ones with the highest, the lowest and median true value (i.e., the resulting expected number of clicks).
To obtain an epistemic distributions for each of our epistemic \acp{PBM}, we sample 2500 parameters from each of them, compute the expected number of clicks conditioned on each sample, and then compute a histogram to approximate the epistemic distributions.
We compare with pointwise predictions which are straightforwardly computed from their learned parameter.

\begin{figure}[t]
	\vspace{0.5\baselineskip}
	\centering
	{
		\renewcommand{\arraystretch}{0.5}
		\setlength\tabcolsep{0.01pt}
		\begin{tabular}{r r r}
			\multicolumn{1}{c}{\footnotesize\;\; Istella \#Feat.\ $34$} &
			\multicolumn{1}{c}{\footnotesize\, Istella \#Feat.\ $172$} &
			\multicolumn{1}{c}{\footnotesize\, Istella \#Feat.\ $195$}
			\\
			\includegraphics[scale=0.45]{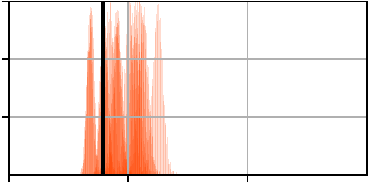}
			&
			\includegraphics[scale=0.45]{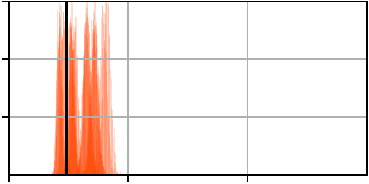}
			&
			\includegraphics[scale=0.45]{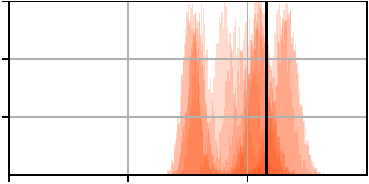}
			\\
			\includegraphics[scale=0.45]{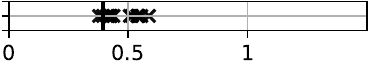}				
			&
			\includegraphics[scale=0.45]{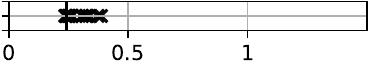}
			&
			\includegraphics[scale=0.45]{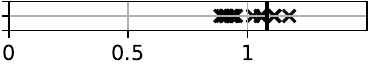}
			\\
			\multicolumn{1}{c}{\footnotesize\;\; MSLR \#Feat.\ $14$} &
			\multicolumn{1}{c}{\footnotesize\, MSLR \#Feat.\ $128$} &
			\multicolumn{1}{c}{\footnotesize\, MSLR \#Feat.\ $133$}
			\\
			\includegraphics[scale=0.45]{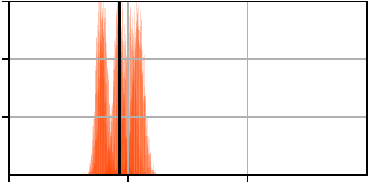}
			&
			\includegraphics[scale=0.45]{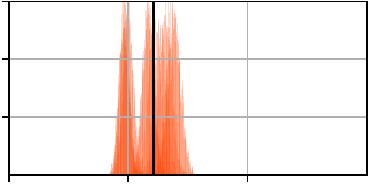}
			&
			\includegraphics[scale=0.45]{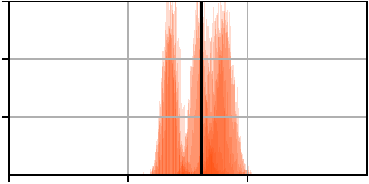}
			\\

			\includegraphics[scale=0.45]{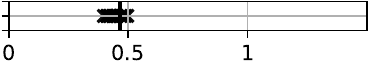}
			&
			\includegraphics[scale=0.45]{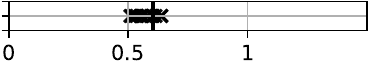}
			&
			\includegraphics[scale=0.45]{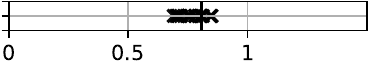}					
		\end{tabular}
	}
	\vspace{-0.5\baselineskip}
	\caption{
		Pointwise and epistemic predictions of the expected number of clicks when using different single features as rankers.
		Vertical lines indicate the ground-truth value.
	}
	\label{fig: Plots Downstream Tasks}
\end{figure}

The results are displayed in Figure~\ref{fig: Plots Downstream Tasks}.
We see that most distributions place substantial probability mass around the ground-truth values, but a few distributions seem to be notably misplaced.
Similarly, the pointwise predictions are also placed around the ground-truth, but are spread out, and thus, many individual pointwise predictions make substantial errors.
The spread and errors of the pointwise and epistemic predictions appear comparable, and are likely a result of the distribution shift.
Yet, unlike the pointwise predictions, the individual epistemic predictions do provide an insight into their possible error through the width of their distributions.

Thus, we affirmatively answer the fifth research question: \emph{Epistemic \acp{PBM} provide useful epistemic predictions for downstream ranker evaluation that appear comparable in accuracy with pointwise predictions while also giving useful indications of uncertainty.}

\section{Conclusion}

We introduced epistemic PBM -- the \emph{first} contextual epistemic click model.
Instead of directly learning \ac{PBM} parameters, our model learns distributions over their possible values.
We optimize with an evidential deep learning approach that maximizes likelihood,
for which we introduced several novel gradient estimation techniques: \emph{self-normalization} and \emph{conditioning on position bias}.
Our experimental results show that \emph{without} these techniques, optimization completely fails; whereas \emph{with} them applied, optimization finds accurate distributions that capture uncertainty appropriately.
Surprisingly, our epistemic \ac{PBM} achieves considerably higher likelihoods than the traditional pointwise \ac{PBM}; an improvement we attribute to the epistemic \ac{PBM}'s capacity for neutral initialization, enabled by the additional confidence parameter. In other words, the epistemic \ac{PBM} allows training to start from uninformative distributions; whereas the pointwise method has no such flexibility and must confidently commit to a random initial estimate, which is almost surely incorrect. This difference is coherently reflected in the initial likelihood: when a prediction is incorrect, low confidence is preferable to a high confidence. Our work has limitations:
Evidential deep learning has known flaws regarding its theoretical points of convergence~\citep{bengs2022pitfalls, pmlr-v235-juergens24a}. %
Furthermore, there is no objective evaluation of epistemic distributions; thus, whilst we did identify some serious failures of baselines, we are limited in assessing the correctness of the shapes of our epistemic distributions.
Nevertheless, we consider this first contextual epistemic click model an important contribution to the field and the start of a new direction of epistemic click modeling research.

\begin{acks}
We want to thank the Institute for Computing and Information Sciences (iCIS) at the Radboud University Nijmegen with which Oscar and Harrie were affiliated while doing the research and writing of this work, in particular, we thank Arjen P. de Vries for his feedback on the manuscript.
This work is supported by the Dutch Research Council (NWO), grant number VI.Veni.222.269 and used the Dutch national e-infrastructure with the support of the SURF Cooperative using grant no.\ EINF-11296. 
\end{acks}

\balance

\bibliographystyle{ACM-Reference-Format}
\bibliography{references}

\end{document}